\documentclass[prc,twocolumn,showpacs,nofootinbib,aps,longbibliography,superscriptaddress]{revtex4-1}
\usepackage[utf8]{inputenc}
\usepackage[T1]{fontenc}
\usepackage{color}
\usepackage{amsmath}
\usepackage{url}
\usepackage{graphicx}
\usepackage{amssymb}
\usepackage{appendix}
\usepackage{multirow}
\usepackage{dcolumn}
\usepackage{hyperref}
\usepackage{mathrsfs} 
\usepackage{bm}
\usepackage{xcolor}
\usepackage{soul}
\usepackage{array}
\usepackage{lipsum}
\usepackage{relsize}
\usepackage{amsfonts}
\usepackage{comment}
\usepackage{outlines}
\usepackage[normalem]{ulem}
\usepackage{natbib}
\usepackage{booktabs}
\usepackage{mathtools}
\usepackage{braket}

\definecolor{linkcolor}{rgb}{0,0,0.40} 
\hypersetup{%
    pdfsubject=Paper,
    pdfkeywords={nuclear physics} {Bayesian} {chiral EFT},
    unicode = true,
    breaklinks = true,
    colorlinks = true,
    linkcolor = linkcolor,
    citecolor = linkcolor,
    menucolor = linkcolor,
    urlcolor = linkcolor
}

\newcommand{\jl}[1]{{\color{blue} #1}}

\newcommand{\vars}{\boldsymbol{\alpha}}

\begin{document}

\title{Emulation of coupled-channels scattering for strongly deformed nuclei}

\author{M. Catacora-Rios}
\email{catacor1@msu.edu}
\affiliation{Facility for Rare Isotope Beams, Michigan State University, East Lansing, Michigan 48824, USA}
\affiliation{Department of Physics and Astronomy, Michigan State University, East Lansing, Michigan 48824, USA}

\author{K. Beyer}
\affiliation{Facility for Rare Isotope Beams, Michigan State University, East Lansing, Michigan 48824, USA}
\affiliation{Department of Physics and Astronomy, Michigan State University, East Lansing, Michigan 48824, USA}

\author{Jin Lei}
\affiliation{School of Physics Science and Engineering, Tongji University, Shanghai 200092, China.}
\affiliation{Southern Center for Nuclear-Science Theory (SCNT), Institute of Modern Physics, Chinese Academy of Sciences, Huizhou 516000, Guangdong Province, China}

\author{F. M. Nunes}
\affiliation{Facility for Rare Isotope Beams, Michigan State University, East Lansing, Michigan 48824, USA}
\affiliation{Department of Physics and Astronomy, Michigan State University, East Lansing, Michigan 48824, USA}

\date{\today}

\begin{abstract}
\begin{description}
\item[Background] Nucleon scattering off strongly deformed nuclei such as the
actinides requires coupled-channels (CC) calculations in which several members
of the ground-state rotational band are coupled explicitly. These calculations
are expensive, which makes Bayesian uncertainty quantification of deformed global
optical potentials costly. A previously developed wavefunction-based CC emulator for inelastic scattering scaled as $\mathcal{O}(N_{c}^{4})$ with the number of channels, because it reconstructs the full matrix of solutions.
\item[Method] We reformulate the direct boundary matching method (DBMM) by imposing its outgoing-wave boundary condition through a generalized
Bloch operator. The resulting Bloch-DBMM solver converges as fast as the modern R-matrix method. We extend the Bloch-DBMM formalism to charged projectiles on targets with
deformed charge distributions. Using this solver we build a reduced-basis emulator,
CCB-ROM, whose basis vectors span all channels of a $J^\pi$ block
simultaneously. We then generalize the coupled-channels empirical
interpolation method (CC-EIM) previously developed to arbitrary non-affine deformed interactions which removes the main computational bottleneck of the previous DBMM-based emulator.
\item[Results] We apply the CCB-ROM emulator to $n,p+{}^{232}$Th inelastic scattering with the ground-state band coupled
through the $8^+$ state, at $4$ and $35$~MeV. The Bloch-DBMM solver is benchmarked against {\sc Frescox}.
The CCB-ROM replaces a $2700$-dimensional system by one of
dimension $N_b\le200$, with a cost of $\mathcal{O}(N_b^3)$ independent of the
number of channels. It reproduces the exact solver across all five band members, including the weakly populated higher-spin states, as well as the transmission coefficients over the full range contributing to the reaction cross section, while retaining speed-ups of over two orders of magnitude relative to Bloch-DBMM. Truncating the basis degrades the weakly populated
channels first, so the accuracy demanded of each state is set by a single
parameter.
\item[Conclusions] CCB-ROM delivers differential cross sections, transmission
coefficients and reaction cross sections from one calculation, at a cost
compatible with Bayesian calibration of deformed optical potentials for
neutrons and protons alike. The Bloch-DBMM solver and the generalized
CC-EIM are applicable to any coupling potential admitting a multipole
decomposition, and the framework extends naturally to closed channels and to
integration with compound-nucleus reaction codes.
\end{description}
\end{abstract}

\maketitle

\section{Introduction}

Reactions on actinides underpin a wide range of nuclear technologies that benefit society \cite{CAPOTE_2008, herman2007empire}. Understanding and predictively modeling these processes is essential for the safe deployment of these technologies. Modeling these reactions typically involves solving a few-body Schr\"odinger equation, where the optical model potential (OMP) plays the role of a complex effective interaction between clusters and carries the nuclear dynamics of the collision.

For actinides, the OMP is obtained from data and then used in compound nucleus codes in two separate steps. The OMP is first calibrated against direct-reaction observables, typically elastic and inelastic angular distributions together with total and reaction cross sections, over a broad energy range. The calibrated potential is then used to compute the transmission coefficients and direct-reaction cross sections that feed statistical and pre-equilibrium reaction codes \cite{KUNIEDA_2007}. Every quantity computed at this second stage therefore inherits the quality of the optical potential and of the parameters determined at the first.

Actinide nuclei are strongly deformed and are typically described as rigid rotors \cite{temura, Buck_1963}. Their low-lying collective states, particularly those of the ground-state rotational band, are strongly excited in nucleon inelastic scattering. The coupling between the elastic and inelastic channels is too strong to treat perturbatively, and a spherical optical model, however carefully calibrated \cite{koning2003local}, cannot reproduce their effect on low-energy cross sections. The Coupled-Channel (CC) formalism, which includes the coupling between these inelastic excitations, is therefore the natural framework for modeling reactions \cite{capote_2005, KUNIEDA_2007}.

 In the standard CC framework, the first several states of the ground-state rotational band are coupled explicitly. This scheme eventually "saturates" in the sense that the higher the excitation of the state in the ground state rotational band \cite{SUKHOVITSKI_2000}, the smaller its impact on the computed observables. Typically, states are included until their contribution is less than the experimental uncertainties. Although tractable, the number of states required for this coupled-channel problem is still large, resulting in roughly $40-100$ partial waves being coupled, depending on the system and the coupling scheme. In addition, coupling schemes beyond the ground state band may be used to refine the reaction description further \cite{Soukhovitski2016}. This gain in fidelity comes at a computational price: calculating reaction observables requires now solving a system of coupled radial equations rather than a set of independent single-channel partial-wave scattering equations. The computational cost of solving such a dense linear system scales cubically with the number of coupled channels. 

Modern nuclear data evaluations increasingly call not only for best-fit cross sections but quantified uncertainties. The preferred framework for uncertainty quantification is Bayesian statistics, which provides full posterior distribution of the OMP parameters, which can be propagated into observables of interest. The parameters of the OMP have been shown to be a major source of uncertainty in the theoretical description of reactions \cite{hebborn2023optical, lovell2017uncertainty, pruitt2023kduq, lovell2020recent,catacora2023complete,catacora2019exploring,catacora2021statistical,beyer2025uncertainty}. A Bayesian uncertainty quantification of this kind requires exploring the parameter space, which in turn demands $10^4$ to $10^6$ evaluations of the scattering observables for each nucleus, every one a full CC calculation repeated across the relevant energies and total spins. For deformed actinides this computational cost has limited Bayesian calibrations to local analyses of individual nuclei. Global parameterizations have been obtained only by frequentist $\chi^2$ fits \cite{KUNIEDA_2007, CAPOTE_2008, soukhovitsky_232Th, capote_2005}. Overcoming this barrier requires a method for evaluating the CC observables at a small fraction of the cost of traditional exact solvers, while preserving their fidelity and precision. This calls for the introduction of emulators, fast and precise replacements for the complex computationally intensive physical model.

Emulators can be classified into two broad classes, those that are physics-driven and those that are data-driven. Reaction  evaluation pipelines require a variety of quantities, including differential and total cross sections, transmission coefficients, etc., so a practical emulator must reproduce all of them reliably and simultaneously, without separate training for each quantity of interest. Therefore, with this interest in mind, we focus on a physics-driven emulator, trained directly on wavefunctions.

Reduced basis methods (RBMs)~\cite{quarteroni2015reducedpde, duguet2024, drischler2023buqeye} provide a wavefunction-based route to emulation.  Rather than solving the full Schr\"odinger equation anew at every parameter value, they solve a projection of it onto a low-dimensional subspace constructed from representative wavefunctions computed across the  physical parameter region of interest.
The reduced equations governing this subspace follow from projecting the scattering operators onto it, using either the Kohn variational principle (KVP)~\cite{kohn1948variational} or a Petrov–Galerkin projection~\cite{quarteroni2015reducedpde, duguet2024}. Because the projection acts on the governing equations themselves rather than on the observables, the resulting emulator retains the structure of the underlying physics while reproducing its predictions at a fraction of the cost, yielding orders-of-magnitude speedups~\cite{furnstahl2020efficient, drischler2021toward, drischler2023buqeye, zhang20213body, melendez2021fast, garcia2023wave, bonilla2022training,odell2024rose}. Recent work on coupled-channels scattering using RBMs demonstrated that an emulator can accelerate neutron–nucleus scattering calculations by factors of $20$–$50$, depending on the target and the number of coupled channels involved \cite{wavefunction-based-catacorarios}. However, the resulting approach, relying on a per-channel RBM expansion, exhibits an unfavorable computational scaling of $\mathcal{O}(N_{\mathrm{ch}}^{4})$, compared to the $\mathcal{O}(N_{\mathrm{ch}}^{3})$ scaling of the exact solver. This limitation arises from the need to emulate the full solution matrix of wavefunctions with linearly independent boundary conditions, which is traditionally required to construct the coupled-channels $\boldsymbol{R}$ matrix and, subsequently, the full scattering matrix $\boldsymbol{S}$ \cite{thompson2009nuclear}. 

In practice, however, the complete $\boldsymbol{S}$ matrix is often not essential for computing experimentally relevant observables or other quantities of interest, such as  transmission coefficients. Only those matrix elements corresponding to the entrance channel, typically the target in its ground state, contribute to the quantities of interest. A recent development by Lei introduced a formulation that efficiently and systematically circumvents the construction of the fundamental matrix of solutions referred to as the Direct Boundary Matching Method (DBMM) \cite{dbmm_lei}. This was done by imposing an outgoing boundary condition in the solution of the scattering equation.  

In the present work, we modify the DBMM approach and derive a Bloch-operator formulation of its scattering boundary condition. The resulting exact solver imposes the same boundary condition as DBMM while recovering the more favorable computational scaling of the modern R-matrix method \cite{ baye2015lagrange, descouvemont2010r, baye_2002, DESCOUVEMONT2016199}. We then build an emulator, which we call Coupled Channels Bloch Reduced Order Model (CCB-ROM), on top of this solver, avoiding the $\mathcal{O}(N_{\mathrm{ch}}^{4})$ scaling of the previous coupled-channels emulator \cite{wavefunction-based-catacorarios}. Bloch-DBMM and the original DBMM impose the same physical outgoing-wave condition, but they differ in how that condition is discretized and in how channel coupling at the boundary is treated. The differences are set out at the end of Sec.~\ref{sec:bloch_dbmm}.

The empirical interpolation method (EIM) was introduced in \cite{odell2024rose} to deal with the optical potential's non-affine parameters, the radius and diffuseness. A key aspect of the present work is the extension of EIM to also handle general non-affine nuclear deformations, an extension required for well deformed nuclei such as actinides, because the Taylor expansion used for small-deformation is no longer valid \cite{wavefunction-based-catacorarios}.  By integrating the newly developed and generalized CC-EIM into the CCB-ROM framework, we eliminate the dominant computational bottleneck (the subspace projection of the coupling potential) identified in the emulator by Lei \cite{emulator_dbmm_lei}. Furthermore, we also generalize the formalism and implement the modified Bloch DBMM approach (Bloch-DBMM) and the associated CCB-ROM to scattering off of a target with long-range Coulomb deformations.

Even for the lightest charged particle, the proton,  Coulomb deformation in well deformed nuclei cannot be neglected when considering inelastic excitation. This is particularly pertinent because many optical potential calibrations rely on $(p,p')$ data. The result of this work is a coupled-channel emulator tailored for nucleon-nucleus scattering off deformed nuclei with large deformation, offering user-defined optimal trade-offs between computational efficiency and accuracy, as is necessary for Bayesian uncertainty quantification analyses, and with the capabilities to predict a large set of observables.

The paper is organized as follows. In Sec.~\ref{sec:formalism} we introduce the coupled-channel formalism starting from the deformed Coulomb extension, the modified CC Bloch-DBMM equations in the Lagrange-Mesh, the RBM projection and the CC-EIM equations that allow us to precompute the non-affine coupling potentials. In Sec.~\ref{sec:Implementation} we outline the implementation and details of the generation of training and solving data sets using the Bloch-DBMM solver. We also include a numerical benchmark of the Bloch-DBMM solver against the  reaction code\jl{~}{\sc Frescox} \cite{fresco}. In Sec.~\ref{sec:Results}, we compare the CCB-ROM results for proton and neutron scattering off of a  deformed nucleus against the high-fidelity Bloch-DBMM solver, analyzing the accuracy, stability, computational scaling and speed gain of the emulator. In Sec.\jl{~}\ref{sec:Conclusions}, we summarize and discuss our findings and the implications of our work. For completeness, we include in Appendix~\ref{app:single_channel_scattering_details} a detailed derivation of the Bloch-DBMM formalism,
and show that the computational scaling of this method is the same as the modern R-matrix.  In Appendix ~\ref{App:coupled-channel-scattering-details}, we discuss the coupled-channels empirical interpolation construction on the Lagrange-Mesh as well as include details on the deformed Coulomb implementation.

\section{Formalism} \label{sec:formalism}
\subsection{Coupled-channel scattering wave equations}

We consider proton scattering from a deformed target whose charge distribution is itself deformed. For an incoming channel $\beta_0$, the deformed Coulomb interaction couples it to other channels through its long-range multipole components, which persist unless the matching is performed at prohibitively large radii. The appropriate way to handle this is by matching to the coupling channel solution.
We organize the solution, for each
$J^\pi$, as
\begin{equation}\label{eq:coupled_channel_wave}
    \psi_{\alpha\beta_0}(r) = e^{i\sigma_{\ell_{\beta_0}}}\big[\,
    \bar{F}_{\alpha}^{(\beta_0)}(r) + \chi_{\alpha\beta_0}(r) \,\big],
\end{equation}
where $\bar{F}_{\alpha}^{(\beta_0)}$ is the coupled-channels regular Coulomb
function, $\chi_{\alpha\beta_0}$ is the coupled-channels scattered wave, and
$\sigma_{\ell_{\beta_0}}$ is the Coulomb phase shift of the entrance channel. The index $\alpha$ runs over the allowed number of coupled channels for each $J^\pi$ block, $\alpha=1, ..., N_{ch}$. Explicitly, $\bar{F}_{\alpha}^{(\beta_0)}$ is the solution of the purely Coulomb coupled system,
\begin{equation}\label{eq:coupled_F}
\begin{aligned}
        \Big[ -\frac{d^2}{dr^2} + \frac{\ell_\alpha(\ell_\alpha+1)}{r^2}
    & - k_\alpha^2 \Big]\bar{F}_{\alpha}^{(\beta_0)}(r)
    \\ & + \sum_{\gamma=1}^{N_{ch}} U^C_{\alpha\gamma}(r)\,\bar{F}_{\gamma}^{(\beta_0)}(r) = 0,
\end{aligned}
\end{equation}
regular at the origin. Since the Coulomb multipoles couple the channels,
Eq.~\eqref{eq:coupled_F} is itself a scattering problem: its regular solution
has an incoming wave only in the entrance channel, but outgoing waves in every
channel. We normalize $\bar{F}^{(\beta_0)}$ so that its incoming part is that
of the free Coulomb wave $F_{\ell_{\beta_0}}$,
\begin{equation}\label{eq:Fbar_asym}
    \bar{F}_{\alpha}^{(\beta_0)}(r) \xrightarrow{r\to\infty}
    \delta_{\alpha\beta_0}F_{\ell_{\beta_0}}(\eta_{\beta_0},k_{\beta_0}r)
    + t^C_{\alpha\beta_0}\,H^+_{\ell_\alpha}(\eta_\alpha,k_\alpha r).
\end{equation}
The Coulomb amplitudes $t^C_{\alpha\beta_0}$ describe Coulomb excitation in the absence
of any short-range interaction; they are an output of the solution, not a
choice. They define the pure-Coulomb S matrix
\begin{equation}\label{eq:SC}
    S^C_{\alpha\beta_0} = \delta_{\alpha\beta_0}
    + 2i\sqrt{\frac{k_\alpha}{k_{\beta_0}}}\,t^C_{\alpha\beta_0},
\end{equation}
which reduces to the identity when the off-diagonal Coulomb coupling vanishes.
The construction of $\bar{F}^{(\beta_0)}$ and $t^C_{\alpha\beta_0}$ from
solutions regular at the origin is given in
Appendix~\ref{App:coupled-channel-scattering-details}.

Here $U^C_{\alpha\gamma}$ is the Coulomb coupling potential, which contains the diagonal point-Coulomb monopole and the multipole couplings, diagonal and off-diagonal, of the full interaction $U_{\alpha\gamma}$. The radial forms of the multipole terms are given by Eq.~\eqref{eq:coulomb_form_factor}. The monopole ($\omega=0$) term of $U^C_{\alpha\gamma}$ takes its point-charge form $\propto 1/r$ everywhere, whereas the higher multipoles ($\omega\ge2$) are the potential of a uniformly charged sphere with a deformed surface, $\propto r^{\omega}$ for $r<R_c$ and $\propto 1/r^{\omega+1}$ for $r>R_c$. We define
\begin{equation}
    \begin{aligned}
        U_{\alpha\gamma} & =(2\mu/\hbar^2)V_{\alpha\gamma} \, ,  \\
        k_\alpha^2 & =(2\mu/\hbar^2)(E-\epsilon_\alpha) \, .
    \end{aligned}
\end{equation}

Substituting Eq.~\eqref{eq:coupled_channel_wave} into the full coupled-channels
Schr\"odinger equation and using Eq.~\eqref{eq:coupled_F}, the scattered wave satisfies the
inhomogeneous coupled system
\begin{equation}\label{eq:scattered_eq}
\begin{aligned}
        \Big[ -\frac{d^2}{dr^2} &+ \frac{\ell_\alpha(\ell_\alpha+1)}{r^2}
    - k_\alpha^2 \Big]\chi_{\alpha\beta_0}(r)
    \\ &+ \sum_{\gamma=1}^{N_{ch}} U_{\alpha\gamma}(r)\,\chi_{\gamma\beta_0}(r)
    = -\sum_{\gamma=1}^{N_{ch}} U^{\mathrm{sr}}_{\alpha\gamma}(r)\,
    \bar{F}_{\gamma}^{(\beta_0)}(r),
\end{aligned}
\end{equation}
where the short-range coupling potential is defined by removing every long range Coulomb
term, diagonal and off-diagonal, from the full interaction,
\begin{equation}\label{eq:sr_pot}
    U^{\mathrm{sr}}_{\alpha\gamma}(r)
    = U_{\alpha\gamma}(r) - U^C_{\alpha\gamma}(r).
\end{equation}
The structure of the source in Eq.~\eqref{eq:scattered_eq} is the central feature of this decomposition. The Coulomb multipole terms of $U_{\alpha\gamma}$ and $U^C_{\alpha\gamma}$ are identical and cancel in $U^{\mathrm{sr}}_{\alpha\gamma}$. The monopole terms differ only inside $R_c$, where $U_{\alpha\gamma}$ has the uniformly charged sphere potential $V^C_{\mathrm{sph}}$ and $U^C_{\alpha\gamma}$ contains point-charge interaction, leaving a finite range correction $\Delta V_C = V^C_{\mathrm{sph}} - Z_pZ_te^2/r$ (see Appendix~\ref{App:coupled-channel-scattering-details}). $U^{\mathrm{sr}}$ is therefore the nuclear interaction plus $\Delta V_C$, and the source in Eq.~\eqref{eq:scattered_eq} vanishes beyond the nuclear surface even though the Coulomb coupling itself is long-ranged. The scattered wave can therefore be solved and matched within a confined region $r<r_0$ not much larger than the range of the nuclear interaction $r_0$, in
contrast to a decomposition built on the uncoupled point-Coulomb functions,
for which the off-diagonal Coulomb multipole couplings with $1/r^{\lambda+1}$ radial dependence and the source term on the right hand side of Eq.~\eqref{eq:scattered_eq} persists to large radii. 

We impose two boundary conditions on the scattered wave. Regularity at the
origin requires
\begin{equation}\label{eq:bc_origin}
    \chi_{\alpha\beta_0}(0) = 0.
\end{equation}
The outer condition follows from the requirement that $\chi_{\alpha\beta_0}$
carry only outgoing flux, the incoming wave being supplied in full by $\bar{F}_{\alpha}^{(\beta_0)}$ . Beyond the range of the nuclear interaction, at our matching radius $r= a$,
Eq.~\eqref{eq:scattered_eq} is homogeneous, and its outgoing
solutions are superpositions of the coupled-channels Coulomb
functions $\bar{H}^+_{\alpha\gamma}$ \cite{thompson2009nuclear},
\begin{equation}\label{eq:asymptotic_scattered_wave}
    \chi_{\alpha\beta_0}(r) = \sum_{\gamma=1}^{N_{ch}}
    [\bar{H}^+(r) ]_{\alpha\gamma}\,t_{\gamma\beta_0}, \qquad r \ge a,
\end{equation}
where $\bar{H}^+_{\alpha\gamma}$ solves Eq.~\eqref{eq:coupled_F} with purely
outgoing behavior, and reduces asymptotically to
$\delta_{\alpha\gamma}H^+_{\ell_\alpha}(\eta_\alpha,k_\alpha r)$. The outgoing nuclear amplitudes are fixed by the scattering matrix through
$t_{\gamma\beta_0} = -\frac{i}{2}\sqrt{\frac{k_{\beta_0}}{k_\gamma}}
\big(S_{\gamma\beta_0} - S^C_{\gamma\beta_0}\big)$,
since the Coulomb-excitation part of the outgoing flux is already carried by
$\bar{F}^{(\beta_0)}$.


Equation~\eqref{eq:asymptotic_scattered_wave} cannot be imposed directly,
since the amplitudes $t_{\gamma\beta_0}$, or equivalently the matrix elements
$S_{\gamma\beta_0}$, are not known in advance. To obtain a condition
involving only $\chi$ and the incoming wave functions, we eliminate the
amplitudes through the logarithmic derivative. Differentiating,
\begin{equation}
    \chi'_{\alpha \beta_0}(r)= \sum_{\gamma=1}^{N_{ch}}
    [\bar{H}^{+\prime}(r) ]_{\alpha\gamma}\,t_{\gamma\beta_0}, \qquad r \ge a,
\end{equation}
the amplitudes drop out at the matching radius,
\begin{equation}\label{eq:bc_outer}
\begin{aligned}
    \chi'_{\alpha\beta_0}(a) &= \sum_{\gamma=1}^{N_{ch}}[\boldsymbol{\gamma}_s]_{\alpha\gamma}\,
    \chi_{\gamma\beta_0}(a), \; \text{where} \\
    [\boldsymbol{\gamma}_s]_{\alpha \gamma} &\equiv \sum_{\delta=1}^{N_{ch}}\big[\bar{H}^{+\prime}(a)\big]_{\alpha\delta}
    \big[\bar{H}^+(a)\big]^{-1}_{\delta\gamma},
\end{aligned}
\end{equation}
and where the prime denotes differentiation with respect to $r$. Once the interior system has been solved for $\chi_{\alpha\beta_0}(a)$, the scattering
S matrix is recovered by inverting the amplitude relation,
\begin{equation}\label{eq:smatrix}
    S_{\alpha\beta_0} = S^C_{\alpha\beta_0}
    + 2i\sqrt{\frac{k_\alpha}{k_{\beta_0}}}\,\sum_{\gamma=1}^{N_{ch}}
    \big[\bar{H}^+(a)\big]^{-1}_{\alpha\gamma}\,
    \chi_{\gamma\beta_0}(a).
\end{equation}
In the absence of short-range interactions the source in
Eq.~\eqref{eq:scattered_eq} vanishes, $\chi\equiv 0$, and
$\boldsymbol{S}=\boldsymbol{S}^C$ reduces to pure Coulomb excitation. When the
off-diagonal Coulomb coupling vanishes, $t^C_{\alpha\beta_0}\to 0$ and
$\boldsymbol{S}^C\to\boldsymbol{1}$.
Here the velocity factor $\sqrt{k_\alpha/k_{\beta_0}}$ normalizes $S$ to unit incoming flux in the entrance channel, so that $\boldsymbol{S}$ is unitary for real potentials. The symmetry of $\boldsymbol{S}$ is a separate property that follows from time-reversal invariance (reciprocity). 

The matrix logarithmic derivative $\boldsymbol{\gamma}_s$ and other matrix equations presented above are the multi-channel generalization of the equations presented in \cite{dbmm_lei}. As previously mentioned this generalization is necessary due to the fact that the coupled outgoing Coulomb functions and their derivatives at the matching radius $r=a$ are in general non-diagonal: the
off-diagonal entries encode the long-range Coulomb mixing of channels which is still
present. We note that when the off-diagonal Coulomb coupling
vanishes  $\bar{\mathbf H}^+$ becomes diagonal and $\boldsymbol{\gamma}_s$
reduces to the familiar diagonal form $[\boldsymbol{\gamma}_s]_{\alpha\alpha}
= k_\alpha H^{+\prime}_{\ell_\alpha}(\eta_\alpha,k_\alpha a)/
H^+_{\ell_\alpha}(\eta_\alpha,k_\alpha a)$, where the prime on $H^+_{\ell_\alpha}$ now denotes differentiation with respect to $\rho=k_\alpha r$. For neutron scattering the asymptotic coupled channel functions reduce to diagonal, charge-independent Riccati-Bessel functions and the formalism above naturally reduces to \cite{dbmm_lei}. In the case for which more than one incoming boundary condition is needed, as in the case for nuclei whose initial state $I_i\neq0$, the calculation must be repeated for each $\beta_0$ allowed.

\subsection{The Bloch-DBMM formulation and the Lagrange-Legendre mesh}\label{sec:bloch_dbmm}
The calculable R-matrix is well established as an efficient, stable, and
complementary solver to traditional finite-difference methods
\cite{baye2015lagrange, descouvemont2010r, DESCOUVEMONT2016199}. For scattering boundary conditions, Lagrange--Legendre functions are a favorable choice to expand the Hamiltonian due to their interpolation properties. A key
ingredient in the modern R-matrix formulation is the Bloch boundary operator, which
symmetrizes the scattering Hamiltonian and ensures smooth matching of the interior solution to its asymptotic form. 
In fact, the DBMM boundary condition imposed on the last row of the Hamiltonian by Lei \cite{dbmm_lei} can be
reformulated using a Bloch operator. This reformulation, which we call
Bloch--DBMM, recovers the favorable basis scaling of the modern R-matrix, as shown in Appendix~\ref{app:single_channel_scattering_details}.
It is therefore convenient to solve Eq.~\eqref{eq:scattered_eq}
by expanding the Hamiltonian in the Lagrange-Legendre basis and using the Bloch operator to enforce
the modified DBMM boundary condition of Eq.~\eqref{eq:bc_outer}. 

Because the boundary condition of Eq.~\eqref{eq:bc_outer} is a matrix relation, the
scalar boundary term $B$ of Bloch operator of the standard R-matrix can no longer enforce it. We
therefore introduce a generalized Bloch operator, which retains the diagonal
derivative in channel space but carries an off-diagonal boundary term $B_{\alpha\gamma}$,
\begin{equation}
    \mathcal{L}(B_{\alpha\gamma})
    = \delta(r-a)\left( \delta_{\alpha\gamma}\frac{d}{dr}
    - \frac{B_{\alpha\gamma}}{r}\right).
\end{equation}
This is the central new ingredient of the Bloch--DBMM formulation: the
off-diagonal entries of $B_{\alpha\gamma}$ contains the long-range Coulomb channel mixing. Adding it to the interior system,
the coupled-channels scattered wave satisfies
\begin{equation}\label{eq:scattered_eq_bloch}
\begin{aligned}
        \Big[ -\frac{d^2}{dr^2} &+ \frac{\ell_\alpha(\ell_\alpha+1)}{r^2}
    - k_\alpha^2 \Big]\chi_{\alpha\beta_0}(r)
    \\ &+ \sum_{\gamma=1}^{N_{ch}} \Big( U_{\alpha\gamma}(r)
    + \mathcal{L}(B_{\alpha\gamma}) \Big)\,
    \chi_{\gamma\beta_0}(r)
    \\ & = -\sum_{\gamma=1}^{N_{ch}} U^{\mathrm{sr}}_{\alpha\gamma}(r)\,
    \bar{F}_{\gamma}^{(\beta_0)}(r).
\end{aligned}
\end{equation}
The condition of Eq.~\eqref{eq:bc_outer} is then recovered by identifying the
boundary term with the matrix logarithmic derivative defined there,
\begin{equation}
    B_{\alpha\gamma} = a\,[\boldsymbol{\gamma}_s]_{\alpha\gamma}
    = a\sum_{\omega=1}^{N_{ch}}\big[\bar{H}^{+\prime}(a)\big]_{\alpha\omega}
    \big[\bar{H}^+(a)\big]^{-1}_{\omega\gamma}.
\end{equation}
If no off-diagonal long-range potential is present, the boundary constant term reduces to its diagonal form,
$B_{\alpha\gamma} \rightarrow a\,\delta_{\alpha\gamma}(\gamma_s)_{\alpha\alpha}
= a\,\delta_{\alpha\gamma}\, k_\alpha H^{+\prime}_{\ell_\alpha}(\eta_\alpha,k_\alpha a)/
H^+_{\ell_\alpha}(\eta_\alpha,k_\alpha a)$. The resulting Hamiltonian with the Bloch-operator is symmetric, the proof is given in Appendix~\ref{App:coupled-channel-scattering-details}.

To solve for the scattered wave, we now express Eq.~\eqref{eq:scattered_eq_bloch}
in the Lagrange--Legendre basis.  As is well known, the property of the Lagrange functions, combined with the Gauss--Legendre quadrature approximation, renders any local potential diagonal on
the mesh, so that $U_{\alpha\gamma}$ contributes only through its values at the
mesh points $r_i = ax_i$, where $x_i$ are the mesh points on the interval $x\in (0,1)$.


In the R-matrix and Bloch-DBMM formalism, each channel in the coupled-channel set is expanded in Lagrange-Legendre functions:
\begin{equation}\label{eq:cc_wave_expansion}
    \chi_{\alpha\beta_0}(r) = \frac{1}{\sqrt{a}}
    \sum_{i=1}^{N} c^{(\beta_0)}_{\alpha i}\,\hat f_i(r/a),
\end{equation}
Projecting Eq.~\eqref{eq:scattered_eq_bloch} onto $\hat f_i(r)$, gives 
linear system of dimensions $(N\cdot N_{ch})\times (N \cdot N_{ch})$:
\begin{equation}\label{eq:cc_linear_system}
    \sum_{\gamma=1}^{N_{ch}}\sum_{j=1}^{N}
    \mathcal{H}_{\alpha i,\,\gamma j}\, c^{(\beta_0)}_{\gamma j}
    = b^{(\beta_0)}_{\alpha i}.
\end{equation}
The exact matrix elements and conventions of the above linear system are expressed in Appendix~\ref{App:coupled-channel-scattering-details}.

Solving Eq.~\eqref{eq:cc_linear_system} gives the coefficients
$c^{(\beta_0)}_{\alpha i}$, from which we can express the wavefunction at the channel radius
$\chi_{\alpha\beta_0}(a) = \sum_i c^{(\beta_0)}_{\alpha i}\hat f_i(1)/\sqrt{a}$. The S-matrix then follows from Eq.~\eqref{eq:smatrix}. It is worth pointing out that since dependence on the incoming channel enters only through the source $b^{(\beta_0)}_{\alpha i}$, the linear system of Eq.~\eqref{eq:cc_linear_system} can be solved for all required incoming $\beta_0$ channels by simply updating the source term appropriately. A discussion of the relationship between Bloch-DBMM and the original DBMM is given in Appendix~\ref{App:coupled-channel-scattering-details}.

\subsection{The reduced-basis equations}

The key component of an RBM emulator is a suitable low-dimensional subspace that
accurately captures the solutions of the scattering equation throughout the desired
physical region of parameter space. Once such a subspace is found, the operators of
Eq.~\eqref{eq:scattered_eq_bloch} can be projected onto it, yielding a reduced
representation of the scattering equation that can be solved for any parameter set
within that region. 
Here, we construct a reduced Bloch-DBMM system, which, like the Bloch-DBMM (and the DBMM formalisms it is based on), but unlike the previous CC-RBM formalism \cite{wavefunction-based-catacorarios}, now explicitly enforces boundary conditions corresponding to the entrance channel $\beta_0$.
This gives it the favorable scaling with the number of coupled channels that the CC-RBM formalism lacked. 
To do this, we find a replacement of the
$(N\cdot N_{ch})\times(N\cdot N_{ch})$ Bloch-DBMM system by a much smaller
$N_b\times N_b$ one.

In the case under consideration, the parameters are those of the deformed optical
potential, which describes the interaction between the nucleon and the target. Let
$\vars = \{\delta_\omega, V_v, r_v, a_v, \dots\}$ denote the set of parameters that
fix its geometry and coupling strengths. Making this dependence explicit, the Hamiltonian linear system reads:
\begin{equation}\label{eq:hamiltonian_explicit_params}
    \boldsymbol{\mathcal{H}}(\vars)\cdot \boldsymbol{c}^{(\beta_0)} = \boldsymbol{b}^{(\beta_0)}(\vars).
\end{equation}

The physical region of parameter space is the range of parameter values for which a solution of the scattering equation is required. For an optical potential this region is finite and specific, since each parameter carries a physical meaning. For example, the radii are set by the size of the target, the diffusenesses by the thickness of its surface, and the depths by the strength of the nucleon--nucleus interaction. A high-fidelity solver such as Bloch-DBMM must return a solution for any parameter set it is given. An emulator need not, and can instead be tailored to this restricted physical subset of parameter space, whose extent defines the training region $\mathcal{T}$.

The reduced representation for the scattering equations we seek can be built by generating $N_s$ \textit{snapshots}, that is, $N_s$
representative solutions $\{\boldsymbol{c}^{(\beta_0)}(\vars_m)\}_{m=1}^{N_s}$ of
the full linear system of Eq.~\eqref{eq:cc_linear_system}, obtained at interaction
parameters $\{\vars_m\}_{m=1}^{N_s} \in \mathcal{T}$. As in
\cite{wavefunction-based-catacorarios, odell2024rose, emulator_dbmm_lei}, we then
perform a Principal Component Analysis (PCA), also known as Proper Orthogonal
Decomposition (POD) \cite{quarteroni2015reducedpde}, to identify the most important
modes of variability of the \textit{snapshots} within $\mathcal{T}$. To do so, the
\textit{snapshots} are arranged as the columns of an $(N\cdot N_{ch})\times N_s$
matrix,
\begin{equation}
    \boldsymbol{\mathcal{C}}^{(\beta_0)} = \big( \boldsymbol{c}^{(\beta_0)}(\vars_1)
    \quad \boldsymbol{c}^{(\beta_0)}(\vars_2) \quad \dots \quad
    \boldsymbol{c}^{(\beta_0)}(\vars_{N_s})\big),
\end{equation}
which is decomposed through the singular value decomposition,
\begin{equation}
    \boldsymbol{\mathcal{C}}^{(\beta_0)} = \boldsymbol{X}^{(\beta_0)}
    \boldsymbol{\Sigma}^{(\beta_0)}\big(\boldsymbol{W}^{(\beta_0)}\big)^\dagger.
\end{equation}
Here $\boldsymbol{X}^{(\beta_0)}$ is the $(N\cdot N_{ch})\times N_s$ matrix whose
columns are the left singular vectors, and which provides an optimal basis for
representing the \textit{snapshots}. The diagonal $N_s\times N_s$ matrix
$\boldsymbol{\Sigma}^{(\beta_0)}$ holds the singular values $\sigma_i$, each
measuring the importance of the corresponding singular vector. Finally,
$\boldsymbol{W}^{(\beta_0)}$ is the $N_s\times N_s$ matrix of right singular
vectors, whose columns give the coordinates of each \textit{snapshot} in that basis.

As in \cite{emulator_dbmm_lei}, the subspace onto which the scattering equation will be projected is obtained by choosing $N_b$ columns from the matrix $\boldsymbol{X}^{(\beta_0)}$,
 \begin{equation}\label{eq:reduced_basis}
     \boldsymbol{X}^{(\beta_0)}_{\text{red}} = \big( \boldsymbol{x}_1^{(\beta_0)} \quad \boldsymbol{x}^{(\beta_0)}_2 \quad ... \quad \boldsymbol{x}^{(\beta_0)}_{N_b}\big ).
 \end{equation}
Note that each column vector $\boldsymbol{x}_n^{(\beta_0)}$ is an $N\cdot N_{ch}$ vector, which is a linear superposition of the \textit{snapshots} and therefore contains all channels in the $J^{\pi}$ block simultaneously. The solution to the linear system of Eq.~\eqref{eq:cc_linear_system} for any new set of interaction parameters $\vars$ within the training region $\mathcal{T}$ can be approximated as a linear combination of these basis elements:
\begin{equation}
    \boldsymbol{c}^{(\beta_0)}(\vars) \approx \boldsymbol{X}_{\text{red}}^{(\beta_0)}\cdot \boldsymbol{d}^{(\beta_0)}(\vars) = \sum_{n=1}^{N_b} d_n^{(\beta_0)}(\vars) \boldsymbol{x}_n^{(\beta_0)}.
\end{equation}
The coefficients $\boldsymbol{d}^{(\beta_0)}(\vars)$ can be obtained by projecting the scattering equation onto the subspace spanned by $\boldsymbol{X}_{\text{red}}^{(\beta_0)}$ and solving this reduced linear system. Explicitly, we define:
\begin{equation}\label{eq:subspace_projection}
\begin{aligned}
     \boldsymbol{\mathcal{H}}_{\text{red}}^{(\beta_0)}(\vars) &\equiv  \big( \boldsymbol{X}_{\text{red}}^{(\beta_0)} \big)^\dagger \boldsymbol{\mathcal{H}}(\vars) \boldsymbol{X}_{\text{red}}^{(\beta_0)} \\
     \boldsymbol{b}^{(\beta_0)}_{\text{red}}(\vars) &\equiv \big( \boldsymbol{X}_{\text{red}}^{(\beta_0)} \big)^\dagger  \boldsymbol{b}^{(\beta_0)}(\vars)
\end{aligned}
\end{equation}
We use the Hermitian projection $\boldsymbol X^\dagger$ in Eq.~\eqref{eq:subspace_projection}. The reduced basis $\boldsymbol X^{(\beta_0)}_{\text{red}}$ is obtained from a singular value decomposition of the complex snapshot matrix and is therefore orthonormal under the Hermitian inner product, $(\boldsymbol X^{(\beta_0)}_{\text{red}})^\dagger\boldsymbol X^{(\beta_0)}_{\text{red}}=\boldsymbol 1$. With this choice, Eq.~\eqref{eq:reduced_system} makes the residual of the full equation orthogonal to the reduced subspace in the same inner product used to construct the basis. The bilinear projection using $\boldsymbol X^T$ would instead preserve the complex symmetry of $\boldsymbol{\mathcal H}$ in the reduced matrix, but we do not exploit that structure.
The resulting reduced system therefore:
\begin{equation}\label{eq:reduced_system}
    \boldsymbol{\mathcal{H}}_{\text{red}}^{(\beta_0)}(\vars) \cdot \boldsymbol{d}^{(\beta_0)}(\vars) =  \boldsymbol{b}^{(\beta_0)}_{\text{red}}(\vars).
\end{equation}
Since each vector $\boldsymbol{x}_n^{(\beta_0)}$ contains all channels simultaneously, the reduced system in Eq.~\eqref{eq:reduced_system} depends only on the number of basis elements chosen $N_b$, not on the number of channels or the number of basis elements per channel used in the Bloch-DBMM formulation. This means the leading order computational cost of solving the reduced system scales as $\mathcal{O}(N_b^3)$, provided the projection of the Hamiltonian and the source term onto the reduced space defined by Eq.~\eqref{eq:subspace_projection} does not need to be repeated for each new parameter set $\vars$. 

Note that the reduced system of Eq.~\eqref{eq:reduced_system} now depends on $\beta_0$.
This system must therefore be solved for each incoming channel $\beta_0$ considered and, as
stated in the previous subsection, for each $J^{\pi}$ block. This requirement
appears to be a general feature of coupled-channel emulation with scattering
boundary conditions. In coupled-channels scattering, the fundamental matrix of
solutions $\hat{Y}_{\alpha\beta}$, or equivalently the full matrix of solutions
$\Psi_{\alpha\beta}$, is an $N_{ch}\times N_{ch}$ object whose columns correspond to the
linearly independent incoming boundary conditions and whose rows correspond to the
outgoing channels \cite{thompson2009nuclear, herman2007empire,
wavefunction-based-catacorarios}. Each column therefore carries a distinct
asymptotic behavior, and the solutions belonging to different $\beta_0$ populate
different regions of the solution space. The left singular vector basis obtained
for $\beta_0=1$ is consequently not an efficient basis for the $\beta_0=2$ problem,
and so on. To our knowledge, this feature is common in all wavefunction-based emulators of coupled-channel scattering that we are aware of \cite{garcia2023wave, wavefunction-based-catacorarios}. It is absent only when a single incoming channel
suffices, as in the barrier penetration and heavy-ion fusion problems of
\cite{hagino2025multichannel, liao2026deformations}, where the emulator is built
for one fixed entrance channel and the penetrability follows from that solution
alone. Should one require the full matrix of solutions, or equivalently the full
scattering matrix $S_{\alpha\beta}$, the reduced system must be solved repeatedly,
once for each boundary condition.

A key advantage of the DBMM formulation by Lei, and therefore of the method
presented here, is that one need only solve the linear system for those incoming
boundary conditions $\beta_0$ that correspond to the physical problem. Once the
scattered wave $\chi_{\alpha\beta_0}$ is obtained, all elements
$S_{\alpha\beta_0}$ for that incoming channel follow simultaneously
from Eq.~\eqref{eq:smatrix}. This resolves the poor scaling identified in
\cite{wavefunction-based-catacorarios}, whose method required the full fundamental
matrix of solutions $Y_{\alpha\beta}$, and circumvents the need to solve for each
$S_{\alpha\beta}$ matrix element separately as in \cite{garcia2023wave}. Typically,
physical problems restrict the target to be initially in its ground state. For
even-even nuclei with a $0^+$ ground state only one $\beta_0$ is then needed, which then yields a computational cost of $\mathcal{O}(N_b^3)$.
In practice, for a target in an initial state $I_i$, at most $2I_i+1$ calculations will be needed.

\subsection{The generalized coupled-channels empirical interpolation method}

The interactions $U_{\alpha\gamma}(r;\vars)$ are in general non-affine in the
parameters $\vars$. For the deformed optical potentials considered in this work, the non-affine parameters considered are: the radius, the diffuseness, and the deformation length. The Hamiltonian matrix can be decomposed as:
\begin{equation}
    \boldsymbol{\mathcal{H}}(\vars) = \boldsymbol{K} + \boldsymbol{V}(\vars).
\end{equation}
The static term $\boldsymbol{K}$ does not change between $\vars$ evaluations, and can therefore be projected to the reduced space once and stored. However, the projection of the interaction term onto the reduced basis,
\begin{equation}\label{eq:V_red_direct}
    \boldsymbol{V}_{\text{red}}^{(\beta_0)}(\vars) \equiv
    \big(\boldsymbol{X}_{\text{red}}^{(\beta_0)}\big)^\dagger\,
    \boldsymbol{V}(\vars)\,\boldsymbol{X}_{\text{red}}^{(\beta_0)},
\end{equation}
cannot be precomputed in general, and would have to be repeated at every new parameter set.
Since $\boldsymbol{V}(\vars)$ lives in the full $(N\cdot N_{ch})$-dimensional
Bloch-DBMM space, this operation can dominate the online cost of the emulator, as found in \cite{emulator_dbmm_lei}.

We remove this cost by extending the CC-EIM  of \cite{wavefunction-based-catacorarios} to handle any deformed interaction on any radial coordinate mesh.
A general coupling potential matrix can be separated into a sum over multipoles, in the Lagrange mesh, this means decomposing the full $(N\cdot N_{ch})\times (N\cdot N_{ch})$ potential $\boldsymbol{V}$ into matrices $\boldsymbol{\mathcal{F}}^\omega$ for each $J^{\pi}$ block:
\begin{equation}\label{eq:V_multipole_matrix}
    \boldsymbol{V}(\vars) = \sum_{\omega=0}^{N_m} \boldsymbol{\mathcal{F}}^{\omega}(\vars),
    \qquad
    \boldsymbol{\mathcal{F}}^{\omega}(\vars)
    = \boldsymbol{\mathcal{V}}^{\omega} \otimes \boldsymbol{f}^{\omega}(\vars),
\end{equation}
where $\boldsymbol{\mathcal{V}}^{\omega}$ is the $N_{ch}\times N_{ch}$ \textit{geometric}
coupling matrix of multipolarity $\omega$, fixed for a given projectile--target
system and coupling scheme, and $\boldsymbol{f}^{\omega}(\vars)$ is the diagonal
$N\times N$ matrix in the Lagrange-Legendre mesh holding the radial form factor at the mesh points. The entire
parameter dependence is carried by $\boldsymbol{f}^{\omega}$.

The CC-EIM expands each radial factor in $N_U$ fixed basis matrices weighted by
parameter-dependent coefficients $e^\omega$, so that
\begin{equation}\label{eq:eim_matrix_main}
    \boldsymbol{\mathcal{F}}^{\omega}(\vars) \approx
    \sum_{m=1}^{N_U} e_{m}^{\omega}(\vars)\,\boldsymbol{G}^{\omega}_{m},
    \qquad
    \boldsymbol{G}^{\omega}_{m} \equiv
    \boldsymbol{\mathcal{V}}^{\omega} \otimes \boldsymbol{u}^{\omega}_{m},
\end{equation}
where $\boldsymbol{u}^{\omega}_{m}$ is the $N\times N$ diagonal matrix holding the $m$-th
interaction basis function on the Lagrange mesh. The construction of this interaction basis
and the determination of the coefficients $e_{m}^{\omega}(\vars)$ are given in
Appendix~\ref{App:coupled-channel-scattering-details}.

Because the matrices $\boldsymbol{G}^{\omega}_{m}$ carry no parameter dependence,
their projection onto the reduced basis is carried out once, in the offline stage,
\begin{equation}\label{eq:eim_projected}
    \boldsymbol{G}^{\omega(\beta_0)}_{m,\,\text{red}} \equiv
    \big(\boldsymbol{X}^{(\beta_0)}_{\text{red}}\big)^\dagger\,
    \boldsymbol{G}^{\omega}_{m}\,\boldsymbol{X}^{(\beta_0)}_{\text{red}},
\end{equation}
each an $N_b\times N_b$ matrix. Eq.~\eqref{eq:V_red_direct} then becomes a
weighted sum of stored matrices,
\begin{equation}\label{eq:V_red_eim}
    \boldsymbol{V}^{(\beta_0)}_{\text{red}}(\vars) \approx
    \sum_{\omega=0}^{N_m}\sum_{m=1}^{N_U}
    e_{m}^{\omega}(\vars)\,\boldsymbol{G}^{\omega(\beta_0)}_{m,\,\text{red}} .
\end{equation}
Only $N_m N_U$ matrices of dimension $N_b$ are involved, and no operation returns
to the high-dimensional Bloch-DBMM space. For each new parameter set, the cost of
building the reduced Hamiltonian is that of evaluating the coefficients
$e_{m}^{\omega}(\vars)$ once for each multipole and summing the terms of Eq.~\eqref{eq:V_red_eim} for each $J^{\pi}$ block, an operation that is independent of both $N$ and $N_{ch}$. The source $\boldsymbol b^{(\beta_0)}(\vars)$ of Eq.~\eqref{eq:cc_source}
is treated in the same way. Here $\bar{\boldsymbol F}^{(\beta_0)}$ denotes
the coupled Coulomb incoming wave of Eq.~\eqref{eq:coupled_F} in the
discretized form of Eq.~\eqref{eq:cc_source}: a vector of length $N N_{ch}$
whose block $\alpha$ holds $\bar F^{(\beta_0)}_\alpha(r)$ expanded in the
Lagrange basis, ordered so that
$\boldsymbol G^{\omega}_{m}=\boldsymbol{\mathcal V}^{\omega}\otimes
\boldsymbol u^{\omega}_{m}$ acts on it channel block by channel block. The
source depends on the parameters only through the nuclear part of
$U^{\mathrm{sr}}$, since $\bar{\boldsymbol F}^{(\beta_0)}$ and the Coulomb
interaction are held fixed. Using the same
expansion as Eq.~\eqref{eq:eim_matrix_main}, the vectors
\begin{equation}\label{eq:src_pillar}
    \boldsymbol s^{\,\omega(\beta_0)}_{m,\text{red}} \equiv
    \big(\boldsymbol X^{(\beta_0)}_{\text{red}}\big)^\dagger\,
    \boldsymbol G^{\omega}_{m}\,\bar{\boldsymbol F}^{(\beta_0)}
\end{equation}
are computed once offline, each of length $N_b$, and the reduced source is
assembled online as
\begin{equation}\label{eq:b_red_eim}
    \boldsymbol b^{(\beta_0)}_{\text{red}}(\vars) =
    -\sum_{\omega=0}^{N_m}\sum_{m=1}^{N_U}
    e^{\omega}_{m}(\vars)\,\boldsymbol s^{\,\omega(\beta_0)}_{m,\text{red}}
    + \boldsymbol b^{C(\beta_0)}_{\text{red}},
\end{equation}
where $\boldsymbol b^{C(\beta_0)}_{\text{red}} =
-(\boldsymbol X^{(\beta_0)}_{\text{red}})^\dagger\,\Delta V_C\,
\bar{\boldsymbol F}^{(\beta_0)}$ is the contribution of the finite-size
Coulomb correction, which is parameter independent and also projected once.
The coefficients $e^{\omega}_{m}(\vars)$ are evaluated once per parameter
set and shared between Eqs.~\eqref{eq:V_red_eim} and \eqref{eq:b_red_eim},
so neither the operator nor the source requires any operation in the full
Bloch-DBMM space during the online stage.


\section{Implementation}\label{sec:Implementation}

The formalism presented in Section~\ref{sec:formalism} is general and can be applied to a wide range of problems provided the coupling potential can be decomposed in multipoles as shown in Appendix~\ref{App:coupled-channel-scattering-details}. In this work, we will focus on an implementation of neutron and proton inelastic scattering off of a $^{232}$Th target, where the collective deformation is modeled as a rigid rotor \cite{temura,thompson2009nuclear}. Several optical potential parametrizations exist for this system, most of which rely on relativistic kinematics and dispersion relations between the optical potential parameters. As a test of our framework, we choose a training region $\mathcal{T}$ centered around the RIPL-3 values of the simplest but still physical non-relativistic, non-dispersive, deformed potentials available, RIPL-3 Ref. 2303 \cite{ignatyuk2004} for $(n,n')$ and  RIPL-3 Ref. 4604 \cite{maslov2004} for $(p,p')$. We emphasize that this choice of parametrization is one of convenience and our CC-EIM framework extends naturally to dispersive interactions. The saturated coupling scheme used in both RIPL-3 potentials include the $0^+,\, 2^+, \, 4^+,\, 6^+,\text{ and } 8^+$ states of the ground state $K^\pi=0^+$ rotational band. In both potential parametrizations the multipoles $\omega=0,\, 2,\, 4$ are included, corresponding to spherical, quadrupole and hexadecapole coupling matrix elements. The parameters $\vars$ of our model contain all short-ranged nuclear parameters, the deformation lengths $\delta_\omega$ for each multipole and the depth, radius and diffuseness of all volume, surface and spin-orbit terms considered. Explicitly:
\begin{equation}
    \begin{aligned}
            \vars = (\, \delta_2,\, \delta_4,\, V,\, r_v, a_v,\, W,\, r_w,\, a_w,\, W_s,\,\\ r_s,\, a_s,\, V_{so},\, r_{so},\, a_{so})
    \end{aligned}
\end{equation}
The nuclear interaction of our model can be written as:
\begin{equation}
    \begin{aligned}
        U(r,\theta;\vars) = & V \tilde f_\text{def}(r,\theta; \boldsymbol{\omega}, r_v, a_v) \\ & + i W \tilde f_\text{def}(r,\theta; \boldsymbol{\omega},  r_w, a_w)\\
        &-i4a_sW_s \frac{d}{dr} \tilde f_\text{def}(r, \theta ; \boldsymbol{\omega}, r_s,a_s) \\
        & + \boldsymbol{\ell}\cdot\boldsymbol{\sigma} \, \left(\frac{\hbar}{m_\pi c}\right)^{2} \frac{1}{r} V_{so}\,\frac{d}{dr}f_\text{WS}(r ; r_{so},a_{so})
    \end{aligned}
\end{equation}
Here $\boldsymbol{\omega}=(\delta_2,\delta_4)$.
Explicitly the deformed radial form factor $\tilde f_\text{def}$ are:
\begin{equation}\label{eq:radial_form_WS}
\begin{aligned}
    \tilde f_\text{def}(r, \theta';\boldsymbol{\omega}, & r_0, a_0) \\& =
    \bigg[1+\text{exp}\Big( \frac{r-r_0A^{1/3} - \sum_{\omega} \delta_\omega Y_\omega^0(\theta')}{a_0}\Big)  \bigg]^{-1}
\end{aligned}
\end{equation}
Note that the spin-orbit term does not include deformation and the radial form $f_{\text{WS}}$ is just a standard Woods-Saxon. Furthermore, in Eq.~\eqref{eq:radial_form_WS} we do not include in the exponent any volume conservation correction term \cite{fresco,thompson2009nuclear}. This is a deliberate choice to preserve simplicity; the implementation is unaffected by its inclusion.

In this work, we do not include in $\vars$ any long-range Coulomb parameters. Specifically we keep $r_C$ fixed at $1.22$ fm, and set $M(E2)=246.0$~$e\,\mathrm{fm}^2$ and $M(E4)=5657.0$~$e\,\mathrm{fm}^4$, extracted from the $\beta_2$ and $\beta_4$ in \cite{maslov2004}. Further details on the deformed Coulomb potential used are shown in Appendix~\ref{App:coupled-channel-scattering-details}. Our implementation of the Coulomb deformation uses a custom Numerov solver to obtain the interior regular CC Coulomb function of Eq.~\eqref{eq:coupled_F}. For the asymptotic CC outgoing Coulomb function $[\bar{H}^+(r) ]_{\alpha\gamma}$, we use a Coulomb propagation scheme which integrates the same Coulomb CC equation inward from $r_\text{asym}$ to the desired channel radius $a$. The Coulomb propagation is done through a modified Numerov implementation extracted from \texttt{Frescox} \cite{fresco} and converted to Python via \texttt{f2py} \cite{peterson2009f2py}. Since the CC Coulomb functions are constant across $\vars$ evaluations, these are computed only once.

The Bloch-DBMM solver and the resulting CCB-ROM emulator are implemented in \texttt{EMBER}, a python code designed for the repeated parameter evaluations required for uncertainty quantification of a deformed optical potential. This implementation relies on an efficient separation of \textit{offline} and \textit{online} computational tasks. In the \textit{offline} stage, all quantities that remain constant across $\vars$ evaluations are generated and stored.

For the Bloch-DBMM solver, these are the matrix elements of the kinetic operator, Eqs.~\eqref{eq:kinetic_offdiag} and \eqref{eq:kinetic_diag}, the Bloch boundary term, Eq.~\eqref{eq:cc_bloch_surface_element}, and the fixed \textit{geometric} coupling matrix elements of the interaction potential, Eq.~\eqref{eq:V_multipole_matrix}. We also store the asymptotic Coulomb functions at the boundary, Eq.~\eqref{eq:smatrix}, the interior coupled-channels regular Coulomb function, Eq.~\eqref{eq:coupled_F}, and the spin-dependent quantities required to construct the cross sections from the $S$ matrix elements. 


In \texttt{EMBER}, the CCB-ROM emulator is built on top of the solver and inherits all of these precomputed quantities. Its additional \textit{offline} tasks are sampling the training space $\mathcal{T}$, solving for the training \textit{snapshots}, performing the POD and generating the reduced basis, Eq.~\eqref{eq:reduced_basis}. The CC-EIM basis is constructed in the same stage, which entails sampling $\mathcal{T}$ and finding the interaction basis functions, Eq.~\eqref{eq:eim_matrix_main}. Once both bases are available, all Bloch-DBMM operators are projected onto the reduced representation, Eqs.~\eqref{eq:subspace_projection} and \eqref{eq:V_red_eim}.

For the Bloch-DBMM solver, the \textit{online} time is the cost of assembling and solving the linear system with the precomputed quantities, Eq.~\eqref{eq:cc_linear_system}. For the CCB-ROM emulator it is the cost of evaluating the CC-EIM coefficients, Eq.~\eqref{eq:EIM_coefficients}, assembling the reduced system, Eq.~\eqref{eq:V_red_eim}, and solving it, Eq.~\eqref{eq:reduced_system}. The \textit{online} speed-up of the emulator relative to the exact Bloch-DBMM solver therefore reflects only the gain from working in the reduced space, and excludes the cost of constructing the bases and precomputed matrices.

\subsection{Details of the Bloch-DBMM solver}

As a validation of our solver, we compare our Bloch-DBMM implementation against the Numerov finite-differences solver \texttt{Frescox}. We consider a $^{232}$Th target and calculate the inelastic scattering of protons and neutrons at two laboratory energies, $E_\text{lab}=4$ MeV and $35$ MeV. The low-energy neutron case is of particular interest because the transmission
coefficients in this region feed the Hauser--Feshbach codes used in nuclear-data evaluations, and reproducing them accurately is a nuclear-data requirement. Furthermore,  low-energy proton scattering off $^{232}$Th sits below the Coulomb barrier and provides a stringent benchmark of our deformed Coulomb implementation, while the high-energy cases test the nuclear part of the calculation.

For the remainder of this work we use a channel radius of $a=18$~fm, and for the proton case we start the Coulomb propagation at $r_\text{asym}=300$~fm, where the off-diagonal Coulomb couplings have fallen below our chosen tolerance of $10^{-7}$. Unless otherwise stated, the Bloch-DBMM solver uses $N=60$ Lagrange--Legendre basis states per channel, the step size of the coupled-channels Coulomb Numerov solvers is fixed at $h=0.005$~fm, and partial waves are included up to $J_\text{max}=20.5$.

The full comparison against \texttt{Frescox} is given in Appendix~\ref{App:coupled-channel-scattering-details}, Fig.~\ref{fig:frescox_vs_bloch} and Tables~\ref{tab:frescox} and \ref{tab:tcoef}. The two codes agree to $10^{-5}$--$10^{-4}$ in the inelastic differential cross sections to each of the five band members, and to $10^{-7}$--$10^{-5}$ in the transmission coefficients and total reaction cross section, which depend only on the moduli of the $S$-matrix elements.


\subsection{Details of the CCB-ROM emulator}

As previously mentioned, in our implementation, the CCB-ROM emulator is built on top of the Bloch-DBMM solver and inherits all of its precomputed quantities. The training space $\mathcal{T}$ is centered on the appropriate RIPL-3 values of the deformed OMP parametrization, with each parameter allowed to vary by $15\%$ about its central value for this initial test. The space is sampled using the \texttt{latinhypercube()} routine of the \texttt{SciPy} library. For the CC-EIM, whose training is inexpensive, we use $n_U=1000$ training points. For the reduced basis we generate $N_s=500$ \textit{snapshots} with the Bloch-DBMM solver, which we found to be the best compromise between training time and emulator accuracy. In general, the number of \textit{snapshots} required depends strongly on the extent of the training space $\mathcal{T}$ and on the desired accuracy of the emulator. 

As shown in Section~\ref{sec:formalism}, each multichannel $J^\pi$ block of the high-dimensional Bloch-DBMM space is projected onto a single reduced linear system of dimension $N_b$. For the higher $J^\pi$ blocks in these $^{232}$Th calculations, the dimension of the problem is typically $\approx 2700$. We consider reduced spaces of dimension $N_b=50$, $100$, $150$, and $200$. Table~\ref{tab:dimensions} compares the size of the emulator's reduced space against that of the full Bloch-DBMM solver, and gives the equivalent number of basis states per channel. This is to be contrasted with the $8$--$16$ per channel required by the previous CC-RBM emulator \cite{wavefunction-based-catacorarios}. For this calculation the emulator computational scaling is of $\mathcal{O}(N_b^3)$ and does not depend explicitly on the number of channels in the CC system. The use of the CC-EIM removes the computational bottleneck identified by \cite{emulator_dbmm_lei}. In our implementation, the CC-EIM and reduced linear systems are solved with the vectorized \texttt{numpy.linalg.solve()} routine. All emulator and solver calculations were performed on the same laptop.

\begin{table}[t]
\begin{ruledtabular}
\begin{tabular}{cccc}
CCB-ROM ($N_b$) & Bloch-DBMM ($N\cdot N_{ch}$) & $N\cdot N_{ch} / N_b$ & $N_b/N_{ch}$ \\
\colrule
 50  & $45\cdot 60 = 2700$ & 54.0 & 1.1 \\
 100 & $45\cdot 60 = 2700$ & 27.0 & 2.2 \\
 150 & $45\cdot 60 = 2700$ & 18.0 & 3.3 \\
 200 & $45\cdot 60 = 2700$ & 13.5 & 4.4 \\
\end{tabular}
\end{ruledtabular}
\caption{\label{tab:dimensions}Dimensions of the coupled-channel equations for
$n,p+{}^{232}$Th with the ground-state band coupled through $8^+$. For the
largest $J^\pi$ blocks, $N_{ch}=45$ partial waves are coupled. In Bloch-DBMM each
channel is expanded in $N=60$ Lagrange--Legendre functions, giving a
high-fidelity system of dimension $N\cdot N_{ch} = 2700$. The CCB-ROM solves a
projected system of dimension $N_b$ in its place; the last column gives the
equivalent number of basis states per channel, $N_b/N_{ch}$.}
\end{table}

All results reported in the next section were obtained on CPU. The linear-algebra formulation of both the Bloch-DBMM solver and the emulator is naturally suited to GPU acceleration: using \texttt{gARC}, an in-house \texttt{JAX} implementation of the modern R-matrix method, we observed speed-ups of over an order of magnitude on GPU relative to CPU. A comparable gain may be expected here, since the operations involved are of the same type, though we have not implemented it and leave a quantitative assessment to future work.

\section{Results}\label{sec:Results}



In coupled-channels calculations on strongly deformed nuclei, the coupling scheme, that is, the states included and the multipole couplings that generate the transitions between them, is set by the model chosen to describe the physics of the collision rather than by the data available to constrain it. States must be added until their effect on the cross section falls below the experimental uncertainties \cite{SUKHOVITSKI_2000}. The resulting coupling scheme is often considerably larger than the set of states for which experimental angular distributions exist.

This sets two requirements for an emulator of inelastic scattering off a strongly deformed nucleus. First, it must reproduce the differential cross section for any member of the coupling scheme, and it must offer the user \textit{tunability}: the choice of how many states to demand accuracy on, and to what degree. Second, it must deliver, from the same calculation, the other quantities entering an evaluation, in particular the total reaction cross section and the transmission coefficients that serve as inputs to compound-nucleus reaction models. The results below show that our emulator meets both requirements across the full rotational band, for neutrons and protons alike, while retaining a speed-up of over two orders of magnitude over the exact Bloch-DBMM solver.

\subsection{Neutrons}

As a first application of the CCB-ROM emulator, we show results for the inelastic scattering of low-energy neutrons at $4$ MeV. This region provides a stringent test of the emulator, since in this low-energy region close to the single-particle resonances, the wavefunction varies rapidly with the parameters. The singular values of the snapshot matrix decay more slowly, and a larger reduced basis is required to reach a given accuracy.

\begin{figure}
    \centering
    \includegraphics[width=1.0\linewidth]{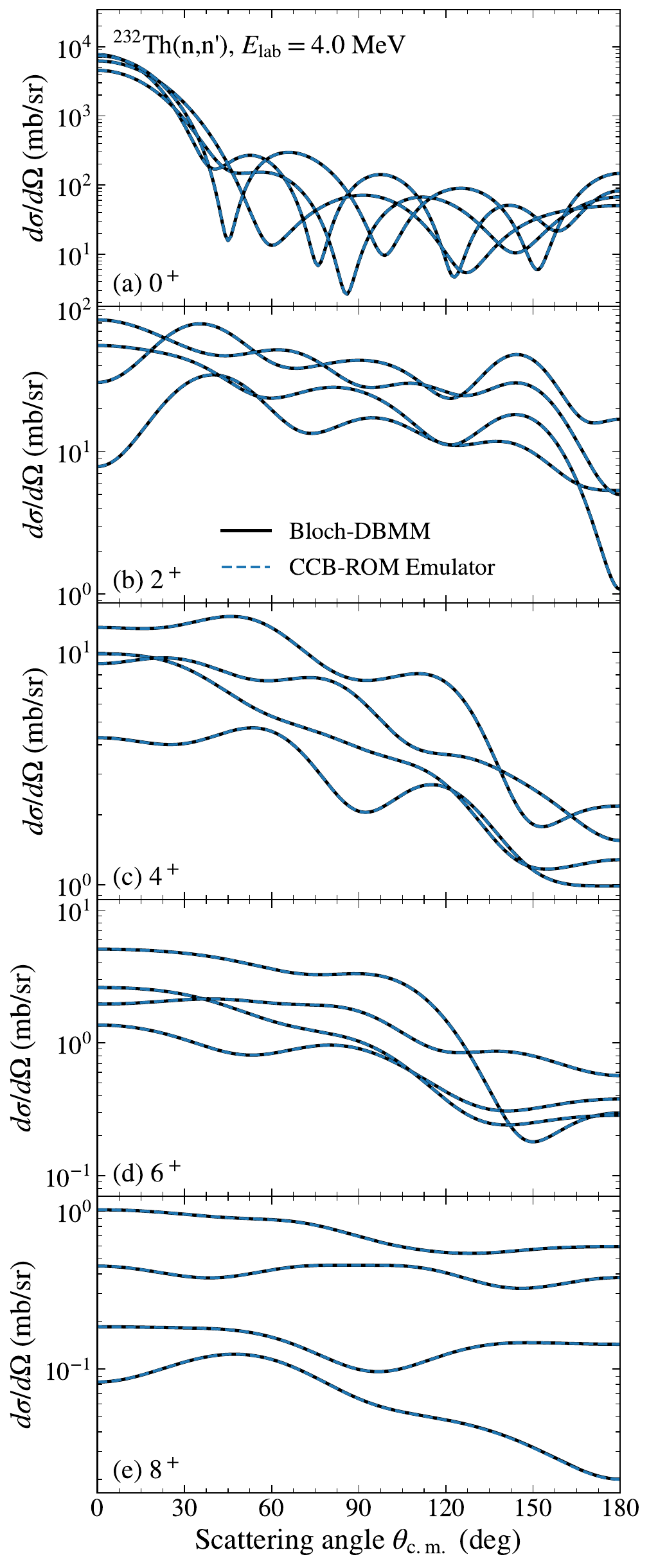}
    \caption{Differential cross sections for $^{232}$Th$(n,n')$ at $E_\mathrm{lab}=4$ MeV, for the five members of the ground-state rotational band: (a) $0^+$, (b) $2^+$, (c) $4^+$, (d) $6^+$, and (e) $8^+$. Solid black lines denote the Bloch-DBMM solver, dashed blue lines the CCB-ROM with $N_b=200$. Each set of curves corresponds to a different parameter set $\vars$ drawn from the training region $\mathcal{T}$. Note the differing vertical scales between panels.}
    \label{fig:low_energy_neutrons_cross_all_bands}
\end{figure}

Figure~\ref{fig:low_energy_neutrons_cross_all_bands} shows the differential cross sections for all five members of the ground-state band, calculated with the emulator at $N_b=200$ and compared against the Bloch-DBMM solver, for four parameter sets drawn from $\mathcal{T}$ not included in the training snapshots. The emulator reproduces the exact result across the full angular range and over the four orders of magnitude separating the $0^+$ and $8^+$ cross sections, including the diffraction structure of the elastic channel and the much flatter angular distributions of the higher band members. The two curves are indistinguishable at the resolution of the figure for every state shown.

\begin{figure}
    \centering
    \includegraphics[width=1.0\linewidth]{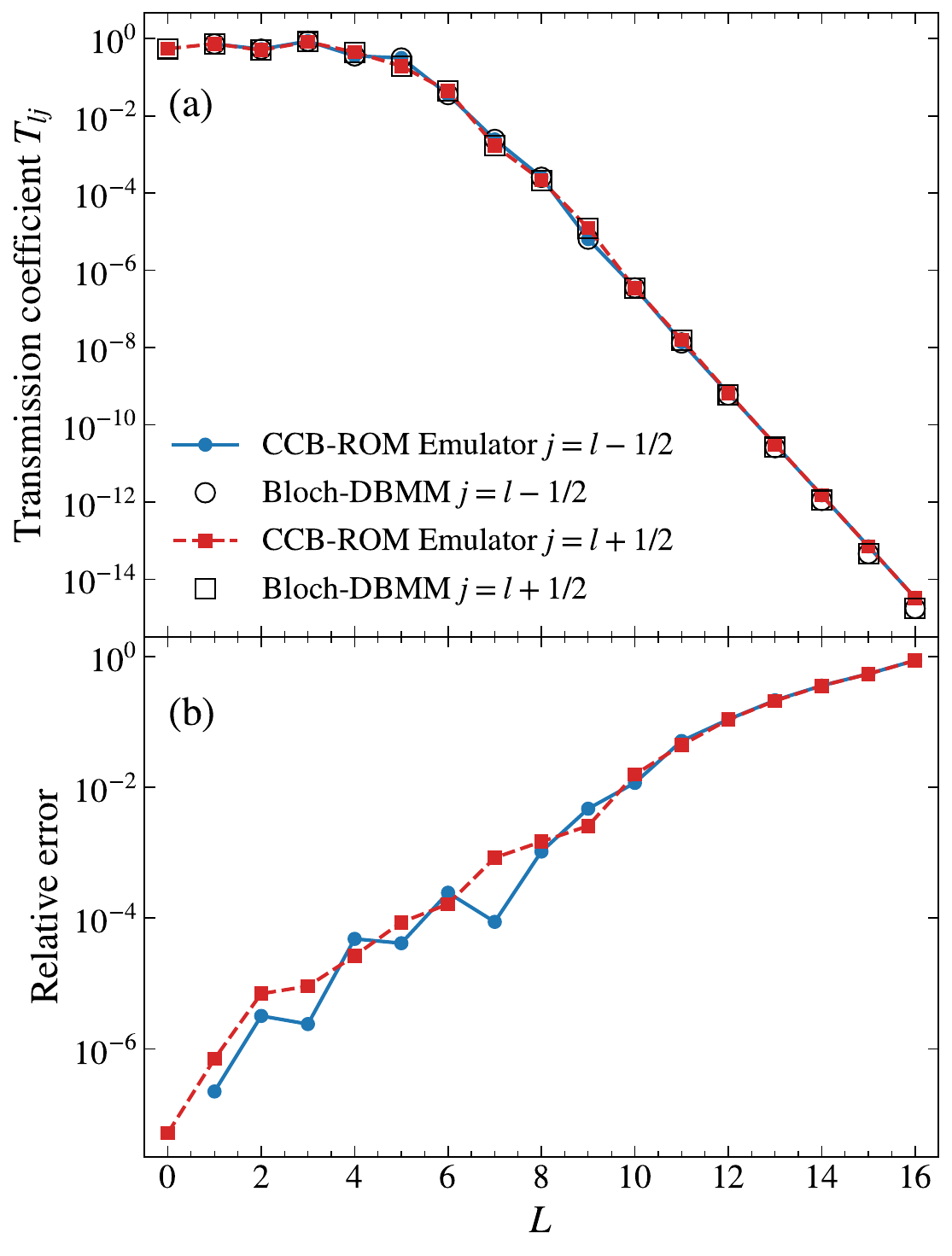}
    \caption{(a) Transmission coefficients $T_{\ell j}$ for $^{232}$Th$(n,n')$ at $E_\mathrm{lab}=4$ MeV, computed with the CCB-ROM emulator (lines) and the Bloch-DBMM solver (open symbols), for $j=\ell-1/2$ (blue circles) and $j=\ell+1/2$ (red squares). (b) Relative error of the emulated coefficients with respect to the Bloch-DBMM result. The emulator calculations were done with $N_b=200$. }
    \label{fig:low_energy_neutrons_transission}
\end{figure}

Figure~\ref{fig:low_energy_neutrons_transission} shows the transmission coefficients $T_{\ell j}$ obtained from the same CCB-ROM calculation, together with their relative error. The entrance-channel transmission coefficients are
\begin{equation}\label{eq:transmission}
    T_{\ell j} \equiv 1 - \sum_{\alpha=1}^{N_{ch}}
    \big|S_{\alpha\beta_0}\big|^{2},
\end{equation}
with $\beta_0$ the entrance channel of angular momenta $(\ell,j)$ and
$S_{\alpha\beta_0}$ from Eq.~\eqref{eq:smatrix}. Because the sum runs over
all channels, $T_{\ell j}$ is the flux absorbed by the imaginary potential,
i.e. the flux into compound-nucleus formation, as required for
Hauser--Feshbach calculations following a coupled-channels calculation. It
is smaller than the total flux leaving the elastic channel,
$1-|S_{\beta_0\beta_0}|^2$, by the direct inelastic probability.
The emulator reproduces the exact coefficients over 7 orders of magnitude in $T_{\ell j}$, with relative errors below $1\%$ for $\ell \le 9$. Beyond this the relative error grows, reaching order unity at $\ell=16$, where $T_{\ell j}$ itself has fallen below $10^{-14}$ and contributes negligibly to any derived quantity. These coefficients are obtained from the same solution as the cross sections, with no separate training, which is
the second requirement set out above.

\begin{figure}
    \centering
    \includegraphics[width=1.0\linewidth]{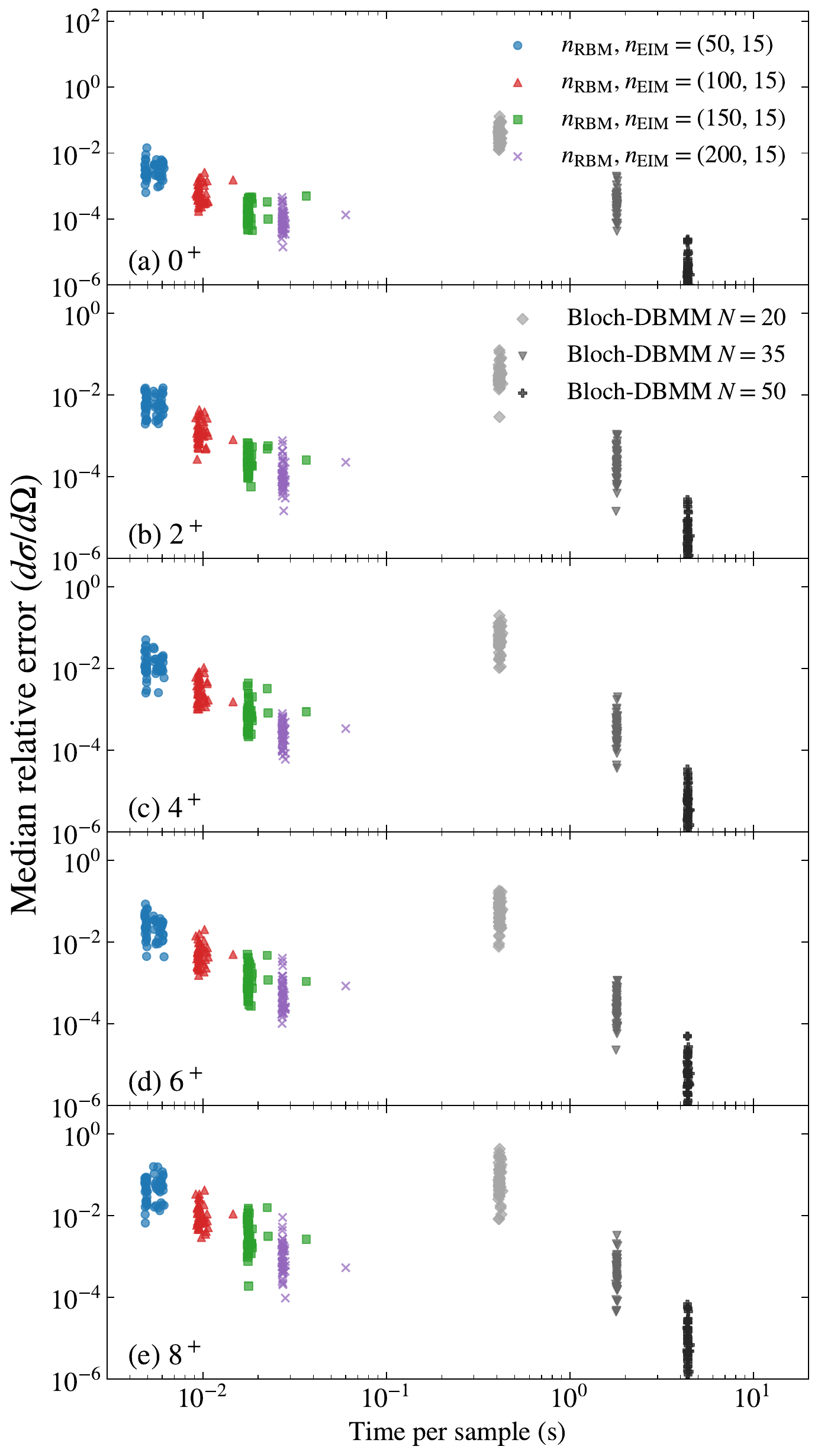}
    \caption{Median relative error in the differential cross section $d\sigma/d\Omega$ against evaluation time per sample, for $^{232}$Th$(n,n')$ at $E_\mathrm{lab}=4$ MeV, shown separately for each member of the ground-state rotational band. Coloured symbols denote the CCB-ROM emulator at reduced-basis dimensions $N_b=50$, $100$, $150$, and $200$, in each case with $N_U=15$ interaction basis functions; grey and black symbols denote the Bloch-DBMM solver at $N=20$, $35$, and $50$ Lagrange--Legendre states per channel. Each point corresponds to one parameter set drawn from $\mathcal{T}$. The median relative error is measured against the $N = 60 $ reference value.}
    \label{fig:low_energy_neutrons_CAT_plot}
\end{figure}

The trade-off between accuracy and evaluation time is quantified in Fig.~\ref{fig:low_energy_neutrons_CAT_plot}, which shows the median relative error per differential cross section ($d\sigma/d\Omega$) calculation across all angles against the time per sample, for reduced bases of dimension $N_b=50$, $100$, $150$ and $200$ and for the Bloch-DBMM solver at $N=20$, $35$ and $50$ Lagrange states per channel. The median relative error is measured against an $N=60$ Bloch-DBMM calculation. Several features are worth noting. First, the emulator error decreases monotonically with $N_b$ for every band member, so the accuracy is controlled by a single parameter and the user may set it as required. Second, at fixed $N_b$, the error grows with band index: at $N_b=50$ the $0^+$ cross section is reproduced to a fractional error of $\approx 3\times10^{-3}$ while the $8^+$ reaches only $\approx 5\times10^{-2}$, and the ordering persists at every basis size. The dominant channels are therefore the last to degrade under truncation, which is a behavior that will be discussed at the end of this subsection.

Comparing the two methods at equal accuracy, the emulator at $N_b=150$ reaches a median error of $\approx 10^{-3}$ in roughly $2\times10^{-2}$ s per sample, against $\approx 2$ s for the Bloch-DBMM solver at $N=35$: a speed-up of two orders of magnitude. At the percent level relevant to reaction calculations, $N_b=100$ suffices for the lower band members and the speed-up is over two orders of magnitude.

We also point out that a harsh truncation of the emulator basis degrades the accuracy of the high spin states ($6^+$ and $8^+$)  while keeping the quality in the description of the lowest states. This follows from each POD basis vector spanning all channels simultaneously: the decomposition is optimal in the norm of the full stacked solution vector, so reducing the basis size first sacrifices the channels carrying the least amplitude. We tested this behavior in the small-deformation limit of the coupling potential, where the cross sections of the higher states are suppressed by seven or more orders of magnitude and correspondingly more basis states are required to reproduce them; the lower angular momentum states retained their accuracy throughout. This \textit{tunability} cannot be reproduced by removing the higher rotational states from the exact calculation, as in a saturated coupling scheme every state is included precisely because its omission shifts the cross sections of all channels significantly.

\subsection{Protons}

We now turn to proton scattering at $E_\text{lab}=35$ MeV. Two features distinguish this case from the $4$ MeV neutron results above. First, the projectile is charged, so the deformed Coulomb coupling is long-ranged and
the channels remain coupled at the matching radius. Second, the beam energy is well above the Coulomb barrier and resonance region, so the wavefunctions vary smoothly with $\vars$ and the snapshot matrix compresses more readily than in the low-energy neutron case.

\begin{figure}
    \centering
    \includegraphics[width=1.0\linewidth]{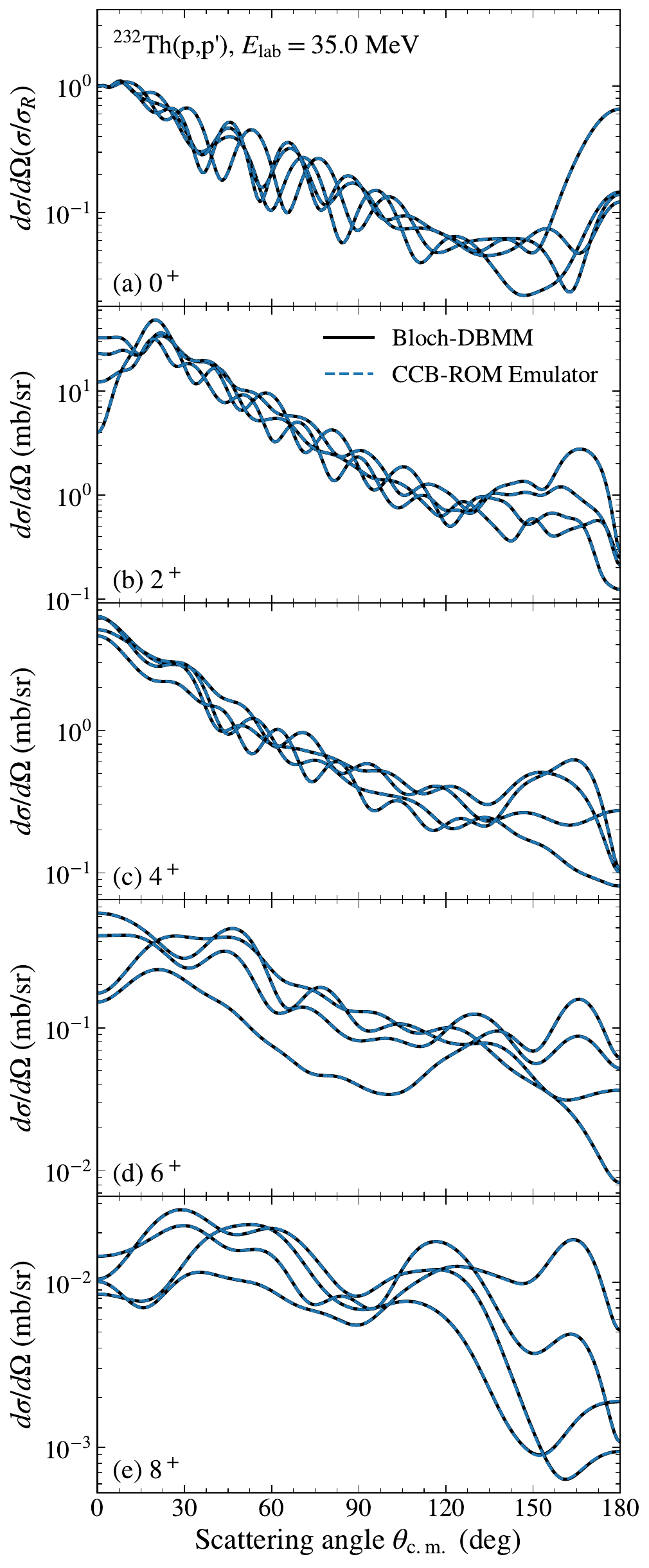}
    \caption{Differential cross sections for $^{232}$Th$(p,p')$ at $E_\mathrm{lab}=35$ MeV, for the five members of the ground-state rotational band: (a) $0^+$, (b) $2^+$, (c) $4^+$, (d) $6^+$, and (e) $8^+$. The elastic cross section in panel (a) is shown as a ratio to the Rutherford cross section; all others are in mb/sr. Solid black lines denote the Bloch-DBMM solver, dashed blue lines the CCB-ROM emulator with $N_b=200$. Each set of curves corresponds to a different parameter set $\vars$ drawn from the training region $\mathcal{T}$. Note the differing vertical scales between panels.}
    \label{fig:proton_xs}
\end{figure}

Figure~\ref{fig:proton_xs} shows the differential cross sections for the five members of the ground-state band. As in the neutron case, the emulator reproduces the exact result across the full angular range, here over the five orders of magnitude separating the elastic ratio from the $8^+$ cross section. The angular distributions are considerably richer than at $4$ MeV, with pronounced diffraction structure persisting into the higher band members, and the emulator follows this structure without degradation.

\begin{figure}
    \centering
    \includegraphics[width=1.0\linewidth]{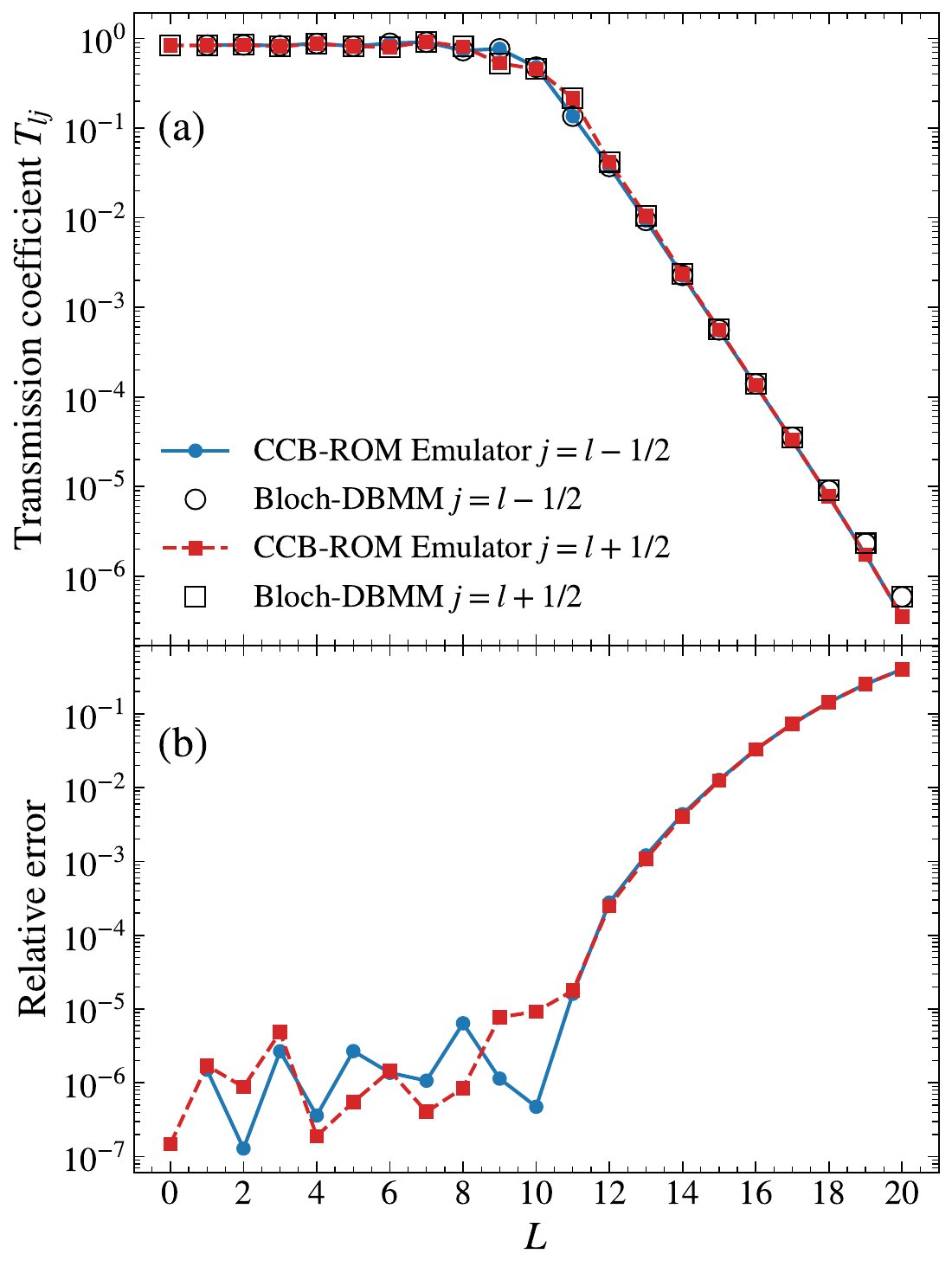}
    \caption{(a) Transmission coefficients $T_{\ell j}$ for $^{232}$Th$(p,p')$ at $E_\mathrm{lab}=35$ MeV, computed with the CCB-ROM emulator (lines) and the Bloch-DBMM solver (open symbols), for $j=\ell-1/2$ (blue circles) and $j=\ell+1/2$ (red squares). (b) Relative error of the emulated coefficients with respect to the Bloch-DBMM result. The emulator calculations were done with $N_b=200$.}
    \label{fig:proton_transmission}
\end{figure}

Figure~\ref{fig:proton_transmission} shows the spin-averaged transmission coefficients and their relative error. The contrast with the neutron case is marked. The Coulomb barrier and the higher beam energy keep $T_{\ell j}$ large out to much higher $\ell$: the coefficients fall by only six orders of magnitude over $\ell=0$--$20$, against fifteen orders at $4$ MeV. 

The relative error correspondingly remains at or below $10^{-5}$ through $\ell = 11$ and below $10^{-2}$ through $\ell = 14$. CCB-ROM therefore reproduces the transmission coefficients to high accuracy across the entire range that contributes appreciably to the reaction cross section.

\begin{figure}
    \centering
    \includegraphics[width=1.0\linewidth]{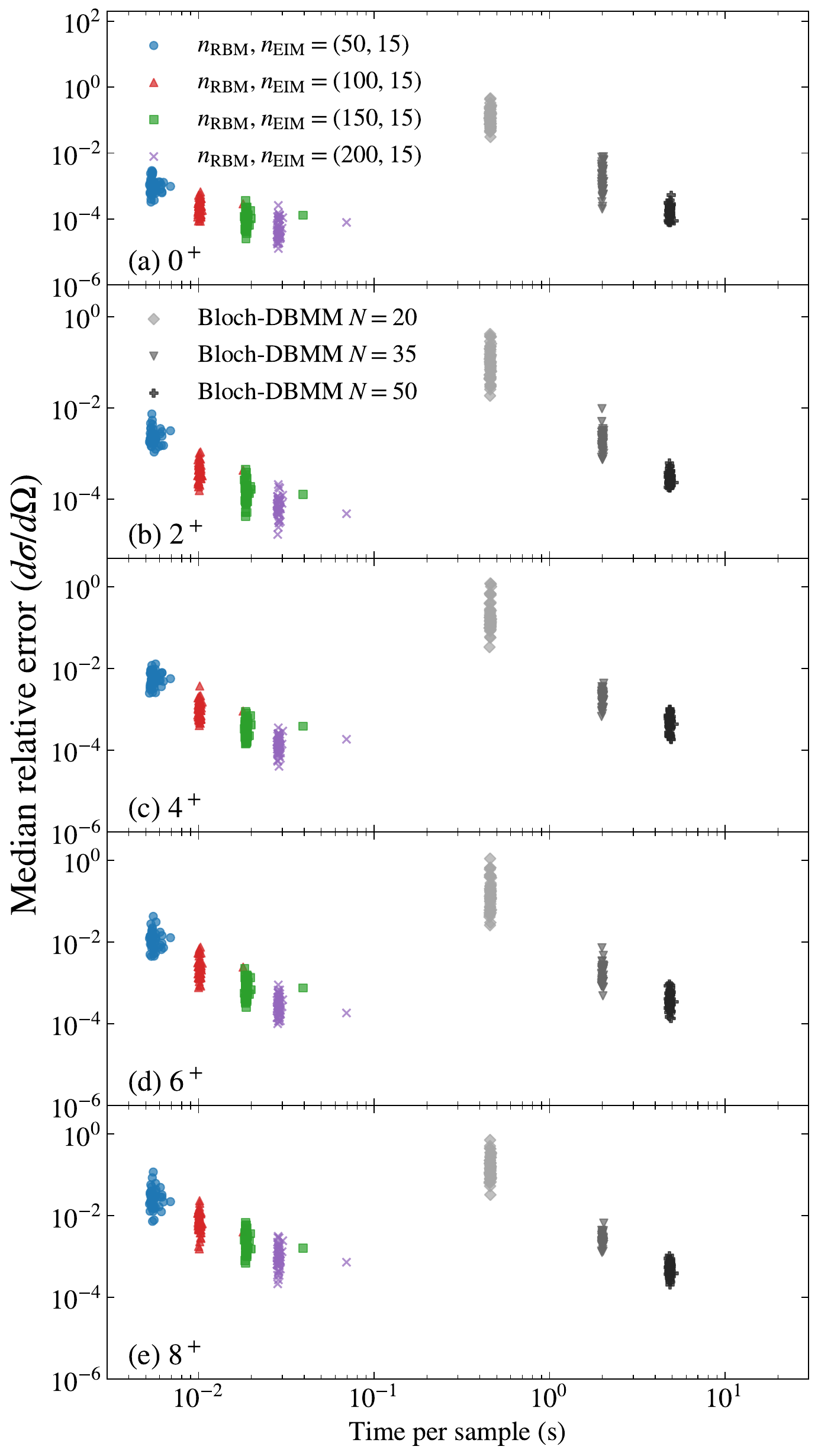}
    \caption{Median relative error in the differential cross section $d\sigma/d\Omega$ against evaluation time per sample, for $^{232}$Th$(p,p')$ at $E_\mathrm{lab}=35$ MeV, shown separately for each member of the ground-state rotational band. Coloured symbols denote the CCB-ROM emulator at reduced-basis dimensions $N_b=50$, $100$, $150$, and $200$, in each case with $N_U=15$ interaction basis functions; grey and black symbols denote the Bloch-DBMM solver at $N=20$, $35$, and $50$ Lagrange--Legendre states per channel. Each point corresponds to one parameter set drawn from $\mathcal{T}$. The median relative error is measured against the $N = 60 $ reference value.}
    \label{fig:proton_cat}
\end{figure}

The accuracy--time trade-off is shown in Fig.~\ref{fig:proton_cat}. The qualitative behavior matches the neutron case: the error decreases monotonically with $N_b$ for every band member, and at fixed $N_b$ it grows for each band member with increasing spin. The comparison against the exact solver, however, differs in an instructive way. At $35$ MeV the Bloch-DBMM calculation is not fully converged at $N=50$: the coupling potential does not vanish within the matching radius of $18$ fm, so resolving it across the box requires more Lagrange--Legendre states than in the neutron case, where the short-range source of Eq.~\eqref{eq:scattered_eq} vanishes well inside $a$. The CCB-ROM emulator does not pay this cost, because the parameters of $\vars$ act only in the nuclear region: beyond the nuclear surface the coupling is purely Coulomb and is identical for every parameter set. The variation the reduced basis must capture is therefore confined to the interior, while the exterior behavior, which the solver must resolve at every evaluation, is carried once by the basis vectors themselves. Constructed from snapshots generated at $N=60$, these vectors inherit the resolution of that calculation, and at $N_b=200$ the emulator matches or exceeds the $N=50$ solver for every band member except the $8^+$. 


Comparing at equal accuracy, the CCB-ROM emulator at $N_b=150$ reaches a fractional median error of $\approx 5\times10^{-4}$ in $2\times10^{-2}$ s per sample, against $\approx 4$ s for the Bloch-DBMM solver at $N=50$, a speed-up of over two orders of magnitude. We note that the additional machinery required by the charged-projectile case carries no online cost to either solver or emulator. Because the Coulomb parameters are not included in $\vars$, the coupled-channels Coulomb functions $\bar{F}^{(\beta_0)}_\alpha$ and $\bar{\mathbf{H}}^+(a)$, and with them the boundary term $B_{\alpha\gamma}$, are computed once and stored in the offline stage; these parameters could in any case be constrained independently through electromagnetic measurements. Including them in $\vars$ would require the regular solution $\bar{F}^{(\beta_0)}_\alpha$ to be obtained by direct Numerov integration at every online evaluation. The evaluation times in Fig.~\ref{fig:proton_cat} are consequently comparable to those of the neutron case, despite the coupled-channel matching at the boundary.

Finally, we note that at energies below the Coulomb barrier, such as the 4 MeV proton case, the cross sections are governed overwhelmingly by the Coulomb interaction, so no emulator is necessary for the cross sections as they are insensitive to the nuclear interaction. 


\section{Conclusions and Outlook}\label{sec:Conclusions}

In this work we have introduced the formalism and framework of CCB-ROM, a wavefunction-based emulator for coupled-channels scattering off heavily deformed nuclei. Because the emulator solves for the scattered wave rather than for a surrogate observable, it predicts differential cross sections, transmission coefficients and the total reaction cross section within the same calculation, with no separate training for each. This is what is required for uncertainty quantification of optical model potentials, and for the emulator to integrate seamlessly into compound-nucleus reaction codes.

The exact solver on which the emulator is built recasts the DBMM boundary condition through a generalized Bloch operator, with the boundary matrix $B=a\boldsymbol\gamma_s$ in place of the $B=0$ of the standard R-matrix method. This symmetrizes the Hamiltonian and recovers the same scaling with basis size as the modern R-matrix, while retaining the central advantage of DBMM: only those incoming boundary conditions corresponding to the physical problem need be solved. We have extended the formalism to targets with deformed charge distributions, where the long-range Coulomb multipoles keep the channels coupled at any practical matching radius and the outer boundary condition becomes a matrix relation. We have also generalized the CC-EIM to arbitrary deformed interactions admitting a multipole decomposition, which removes the projection bottleneck identified in Ref.~\cite{emulator_dbmm_lei} and makes the online cost independent of the number of Lagrange mesh points.

Benchmarking against {\sc Frescox} for $n,p+{}^{232}$Th with the ground-state band coupled through $8^+$, the solver agrees at the $10^{-5}$--$10^{-4}$ level in the differential cross sections and at $10^{-7}$--$10^{-5}$ in the transmission coefficients and reaction cross section. The emulator reproduces the exact solver across all five band members and the four to five orders of magnitude separating the elastic and $8^+$ cross sections, at speed-ups of two orders of magnitude depending on the accuracy demanded. Because each reduced basis vector spans all channels of a $J^\pi$ block simultaneously, the dimension of the reduced system is set by $N_b$ alone: for $^{232}$Th the cost no longer depends on the number of coupled channels and scales as $\mathcal{O}(N_b^3)$, replacing a $2700$-dimensional system by one of dimension $200$ for the largest blocks. We note that a separate reduced basis is required for each incoming channel $\beta_0$, so targets demanding more than one entrance channel, such as those whose initial state $I_i\neq0$, carry a proportionally larger cost. Even in that case the gain over the exact solver remains substantial, since the same scaling with $N_b$ applies to each.

The same property gives the CCB-ROM emulator its \textit{tunability}. Truncating the basis degrades the weakly populated channels first, leaving the cross-section dominant states accurate, so the accuracy demanded of each band member is set through a single parameter and can be determined by how many states are available in the dataset. This cannot be reproduced by truncating the coupling scheme of the exact solver itself, since in a saturated coupling scheme removing any state alters all remaining cross sections significantly.

We note that although the reduced linear system is small, a significant part of the evaluation time reported in the accuracy-versus-time (CAT) plots of Figs.~\ref{fig:low_energy_neutrons_CAT_plot} and \ref{fig:proton_cat} is not spent solving it. Assembling the observables from the emulated solution, which requires moving the precomputed spin coefficients and boundary quantities through memory, accounts for a substantial fraction of the online cost. Further gains in this framework are therefore to be sought in the construction of the observables rather than in reducing $N_b$, and this is where GPU execution, natural to the linear-algebra formulation, is likely to help most.

We also find that the required basis size depends on the scattering energy in a way not captured by the reduced dimension alone. At $4$ MeV, where single-particle structure lies close to the scattering energy, the wavefunctions vary sharply across the training region and the singular values of the snapshot matrix decay more slowly than at $35$ MeV. In applications spanning a broad energy range, we need to adjust the basis size required for a given accuracy accordingly.

Several extensions follow naturally. The inclusion of closed channels is the most immediate: at low incident energies these contribute significantly to the cross sections, and their treatment within the Bloch-DBMM formulation requires only a change in the asymptotic form imposed at the channel radius. With the increased speed in the calculations of the direct reactions part, it would be of interest to incorporate our method into the existing compound nucleus codes such as \texttt{TALYS} \cite{talys} or \texttt{YAHFC} \cite{yahfc}, thus being able to do calibration and uncertainty quantification of both direct and compound nucleus model parameters simultaneously.

\section*{Acknowledgements}
We thank A. Sieverding, O. C. Gorton, P. McGlynn, P. G. Giuliani, C. D. Pruitt and I. J. Thompson for many valuable discussions, and C. Hebborn for a careful reading of the manuscript and valuable comments and suggestions. This research was supported by the CSSI program Award OAC-2004601 (BAND collaboration). F.M.N. and K.B. acknowledge  support of the U.S. Department of Energy Grant No. DE-SC0021422. J.L. acknowledges support from the National Natural Science Foundation of China (Grants No. 12475132 and No. 12535009).

\clearpage

\appendix

\section{Single-Channel Scattering }\label{app:single_channel_scattering_details}
\subsection{Derivation of the boundary condition imposed via the Bloch operator} 
Let us consider the case of single-channel elastic scattering to demonstrate how the boundary condition of \cite{dbmm_lei} can be imposed as a Bloch operator. In the single-channel case the wavefunction of the scattering process after partial wave expansion can be expressed as:
\begin{equation}\label{eq:single_channel_wave}
    \psi_l(r) = F_l(\eta, kr) + \chi_l(r).
\end{equation}
The single-channel Hamiltonian is
\begin{equation}\label{eq:single_channel_hamiltonian}
   H(r) =  -\frac{d^2}{dr^2} + \frac{l(l+1)}{r^2} + \frac{2\mu}{\hbar^2} U(r),
\end{equation}
with $U(r)=V_N(r)+V_c^s(r)+U_c(r)$ the sum of the short-range nuclear and Coulomb potentials and the long-range Coulomb potential. The regular Coulomb function satisfies $[-d^2/dr^2+l(l+1)/r^2+U_c(r)-k^2]F_l(\eta,kr)=0$, so acting with $H$ on the wavefunction of Eq.~\eqref{eq:single_channel_wave} casts the
problem in inhomogeneous form,
\begin{equation}
\label{eq:single_channel_inhomogeneous}
-\chi_l''(r) + Q_l(r)\,\chi_l(r) = s_l(r),
\end{equation}
with
\begin{equation}
\begin{aligned}
Q_l(r) &\equiv \frac{l(l+1)}{r^2} + \frac{2\mu}{\hbar^2}\,U(r) - k^2, \\
s_l(r) &\equiv -\frac{2\mu}{\hbar^2}\big(V_N(r)+V_c^s(r)\big)\,F_l(\eta,kr).
\end{aligned}
\end{equation}
We impose on $\chi_l$ the outgoing-wave boundary condition in
logarithmic-derivative form at $r=a$,
\begin{equation}\label{eq:boundary_condition_explicit}
    \chi_l'(a) = \gamma_l\,\chi_l(a),
\end{equation}
with $\gamma_l = k\,H_l^{'+}(\eta,ka)/H_l^+(\eta,ka)$, where the prime on $H^+_l$ denotes differentiation with respect to $\rho=kr$, from which the T-matrix $\boldsymbol{T}_l^{\text{scat}}=\chi_l(a)/[kH_l^+(\eta,ka)]$ and
\begin{equation}\label{eq:s_matrix_single_channel}
    S_l = 1+2ik\,\boldsymbol{T}_l^{\text{scat}}
\end{equation}
follow.

In \cite{dbmm_lei} this condition is imposed at the last point of the Lagrange--Legendre mesh. We instead impose it through a Bloch operator, which symmetrizes the Hamiltonian and recovers the computational convergence of the R-matrix. Taking the weak form of Eq.~\eqref{eq:single_channel_inhomogeneous} --- projecting onto any $w(r)$ regular on $[0,a]$ with $w(0)=0$, as in the Galerkin construction of several RBM emulators \cite{odell2024rose,bonilla2022training} --- and integrating the second-derivative term by parts gives
\begin{equation}
\label{eq:weak_form_boundary}
\begin{split}
&\int_0^a w'(r)\,\chi_l'(r)\,dr - \big[\,w(r)\,\chi_l'(r)\,\big]_0^a \\
&\quad + \int_0^a w(r)\,Q_l(r)\,\chi_l(r)\,dr = \int_0^a w(r)\,s_l(r)\,dr .
\end{split}
\end{equation}
The surface term vanishes at $r=0$ because $w(0)=0$; imposing the desired boundary condition
Eq.~\eqref{eq:boundary_condition_explicit} at $r=a$ then yields
\begin{equation}
\label{eq:weak_form_final}
\begin{split}
&\int_0^a w'(r)\,\chi_l'(r)\,dr - w(a)\,\gamma_l\,\chi_l(a) \\
&\quad + \int_0^a w(r)\,Q_l(r)\,\chi_l(r)\,dr = \int_0^a w(r)\,s_l(r)\,dr ,
\end{split}
\end{equation}
whose left-hand side is manifestly symmetric under the interchange $w\leftrightarrow\chi_l$.

This is precisely the effect of augmenting the Hamiltonian with the Bloch
operator~\cite{BLOCH1957503}
\begin{equation}
\label{eq:bloch_operator}
\mathcal{L}(B) = \delta(r-a)\left(\frac{d}{dr} - \frac{B}{r}\right),
\qquad B = a\gamma_l,
\end{equation}
giving the symmetrized Hamiltonian
\begin{equation}
\label{eq:single_channel_hamiltonian_bloch}
\mathcal{H}(r) = -\frac{d^2}{dr^2} + \frac{l(l+1)}{r^2}
+ \frac{2\mu}{\hbar^2}\,U(r) + \mathcal{L}(a\gamma_l) .
\end{equation}
Since $\mathcal{L}(a\gamma_l)$ has support only at $r=a$, its projection is
\begin{equation}
\label{eq:bloch_surface}
\int_0^a w(r)\,\mathcal{L}(a\gamma_l)\,\chi_l(r)\,dr
= w(a)\big[\chi_l'(a) - \gamma_l\,\chi_l(a)\big] .
\end{equation}
Projecting $[\mathcal{H}(r)-k^2]\chi_l(r)=s_l(r)$ onto $w(r)$ and integrating the kinetic term by parts, the surface contribution $w(a)\,\chi_l'(a)$ cancels the corresponding term in Eq.~\eqref{eq:bloch_surface}, reproducing Eq.~\eqref{eq:weak_form_final} exactly. The explicit surface term and the Bloch operator with $B=a\gamma_l$ are therefore interchangeable. To solve this system we can expand the scattered wave in the Lagrange--Legendre basis,
\begin{equation}\label{eq:scat_wave_lagrange_basis_expansion}
    \chi_l(r) = \frac{1}{\sqrt{a}}\sum_{j=1}^N c_j\,\hat{f}_j(x),
\end{equation}
with $x=r/a$ and mesh points $r_i=ax_i$. Projecting
$[\mathcal{H}(r)-k^2]\chi_l(r)=s_l(r)$ onto $w(r)=\frac{1}{\sqrt{a}}\hat{f}_i(x)$
and using the Gauss--Lagrange property that local operators are diagonal on the
mesh gives
\begin{equation}\label{eq:single_channel_hamiltonian_LL_basis}
\begin{split}
    \mathcal{H}_{ij}(a\gamma_l) = {}& M_{ij}(a\gamma_l) \\
    & + \bigg[\frac{l(l+1)}{r_i^2} + \frac{2\mu}{\hbar^2}U(r_i)-k^2\bigg]\delta_{ij},
\end{split}
\end{equation}
where, with $T=-d^2/dr^2$, the kinetic-plus-Bloch matrix elements are
\begin{equation}\label{eq:single_channel_kinetic_matrix}
\begin{split}
    M_{ij}(a\gamma_l) = {}& \langle \hat{f}_i | T + \mathcal{L}(0) | \hat{f}_j \rangle \\
    & - \Big\langle \hat{f}_i \Big| \delta(r-a)\frac{a\gamma_l}{r} \Big| \hat{f}_j \Big\rangle .
\end{split}
\end{equation}
The first term is the symmetric R-matrix kinetic matrix; the second carries the
boundary condition through $a\gamma_l$. Next, we discretize the source of
Eq.~\eqref{eq:single_channel_inhomogeneous} on the same mesh,
\begin{equation}\label{eq:source_discrete}
    s_i = -\frac{2\mu}{\hbar^2}\big(V_N(r_i)+V_c^s(r_i)\big)\,F_l(\eta,kr_i)\,\sqrt{a\lambda_i},
\end{equation}
with $\lambda_i$ the Gauss--Legendre weights. The coefficients follow from solving the
linear system
\begin{equation}\label{eq:linear_system}
    \mathcal{H}_{ij}(a\gamma_l)\,c_j = s_i .
\end{equation}
This is exactly the system solved by existing R-matrix codes \cite{descouvemont2010r}; the only additions
are the Bloch boundary term in Eq.~\eqref{eq:single_channel_kinetic_matrix} and
the source $s_i$, which makes the formulation straightforward to implement.

\subsection{Convergence of the Bloch-DBMM exact solver}

In the canonical DBMM reference \cite{dbmm_lei}, the convergence of the exact
DBMM solver with the number of basis states was found to be slower than that of the modern R-matrix. Our implementation of the original DBMM solver showed that one needs roughly 1.5 to 2 times the number of basis states to reach the same accuracy, in direct agreement with the results presented in \cite{dbmm_lei}. One possible reason, identified in that reference, is the variational principle of the R-matrix method \cite{lane1958r}. In practice, the Bloch operator in the R-matrix formulation symmetrizes the Hamiltonian, which leads to more stability and better convergence. By contrast, DBMM discretizes the strong form of the equation by collocation. The non-symmetric derivative matrix acts at the $N-1$ interior mesh points, and the equation at the last point is replaced by the boundary condition. The resulting linear system is not the projection of a symmetric operator, so the weak form of Eq.~\eqref{eq:weak_form_boundary} never appears. As shown in the previous subsection, by imposing the same boundary condition as DBMM but through the Bloch operator, we symmetrize the Hamiltonian and recover the same convergence as the R-matrix. For the large coupled-channel systems considered in this work, the resulting speedup in training with the Bloch--DBMM solver is significant.

\begin{table}
\begin{ruledtabular}
\begin{tabular}{ccc}
$N$ & Bloch--DBMM & R-matrix \\
\colrule
15 & $1.34\times10^{-1}$ & $2.14\times10^{-1}$ \\
30 & $1.05\times10^{-3}$ & $1.35\times10^{-3}$ \\
40 & $3.25\times10^{-5}$ & $4.05\times10^{-5}$ \\
50 & $6.95\times10^{-7}$ & $1.10\times10^{-6}$ \\
\end{tabular}
\end{ruledtabular}
\caption{\label{tab:convergence}Median relative error in the differential cross
section $d\sigma/d\Omega$ with respect to the high-tolerance Runge--Kutta
reference, versus the number of Lagrange--Legendre basis states $N$, for the
Bloch--DBMM and modern R-matrix methods. System: $n+{}^{40}$Ca at 20~MeV,
channel radius $a=20$~fm.}
\end{table}

We now compare the convergence of the modern R-matrix and the Bloch--DBMM solver
as a function of the number of basis states. For simplicity we focus on
single-channel elastic scattering, in particular $n+{}^{40}$Ca at 20~MeV with
the channel radius set to $a=20$~fm. The interaction comprises real and
imaginary volume terms together with an imaginary surface term, with all
Woods--Saxon parameters set to the appropriate Koning--Delaroche values
\cite{koning2003local} taken from \cite{capote2009ripl}. We track the
convergence of the elastic cross section, computed with 20 partial waves,
against a Python-based Runge--Kutta reference with tolerances set to $10^{-12}$.
As Table~\ref{tab:convergence} shows, the Bloch--DBMM solver matches or exceeds
the convergence of the R-matrix for this case, with both methods reaching
sub-percent accuracy at around 30 Lagrange--Legendre basis functions.

\begin{figure}
    \centering
    \includegraphics[width=1.\linewidth]{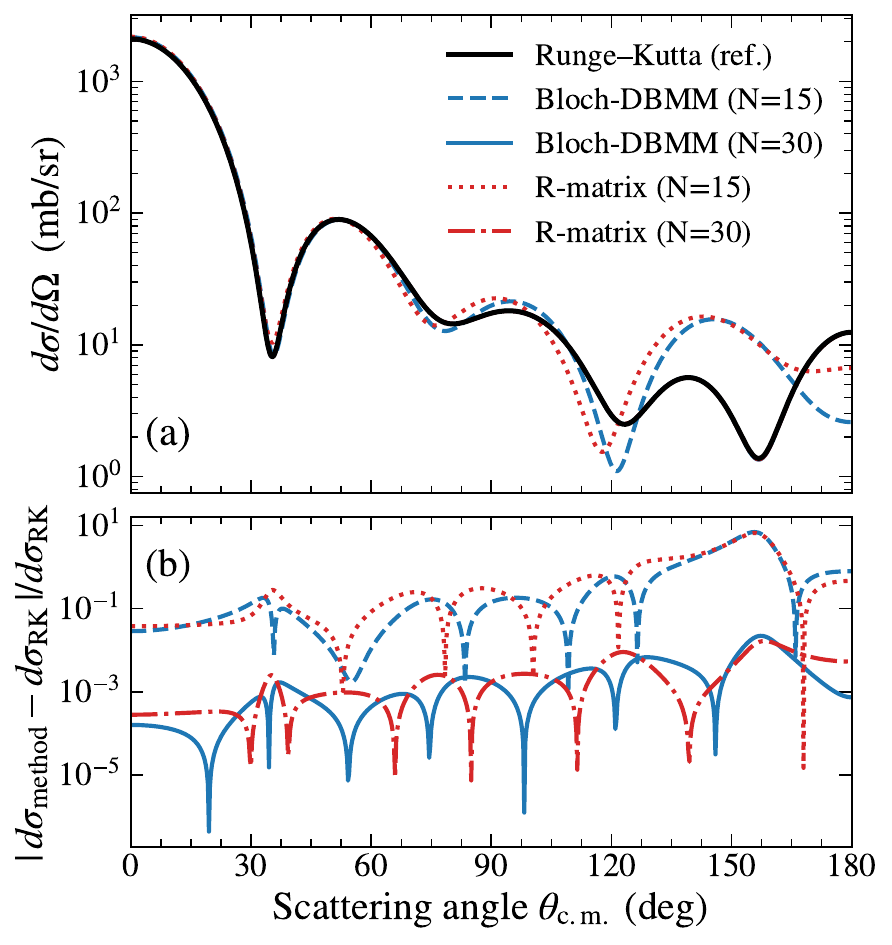}
    \caption{Convergence of the R-matrix and Bloch-DBMM solvers as a function of the umber of Lagrange--Legendre basis states, benchmarked against a high-tolerance Runge--Kutta reference, for $n+{}^{40}$Ca at 20~MeV. Top: differential cross section; bottom: percent relative error at each angle. Bloch--DBMM is shown in blue ($N=15$ dashed, $N=30$ solid) and the modern R-matrix in red ($N=15$ dotted, $N=30$ dash-dotted), with the Runge--Kutta reference in solid black. Both methods reach sub-percent accuracy at $N=30$.
The channel radius is $a=20$~fm.}
    \label{fig:jitr_v_dbmm_nbasis}
\end{figure}

In Figure~\ref{fig:jitr_v_dbmm_nbasis} we show the differential cross section and
relative error for the $n+{}^{40}$Ca system at several basis sizes. The top
panel displays the cross section computed with $N=15$ and $N=30$ basis states
for both methods, together with the Runge--Kutta reference; the bottom panel
displays the corresponding percent relative error at each angle. At $N=15$ both
methods deviate visibly from the reference, whereas at $N=30$ both curves lie on top of the reference
and the relative error remains below one percent across the full angular range.
In the other systems and energies we examined, including cases beyond
single-channel scattering, we observed a similar convergence behavior between the
two methods. For the coupled-channel calculations considered in this work, a comparison between Bloch-DBMM and the R-matrix solver reached median relative errors of $ \approx 10^{-5} $ to $10^{-8}$ in the cross section for fully converged calculations with the same number of Lagrange-Legendre basis states.

\section{Coupled-Channel Scattering}\label{App:coupled-channel-scattering-details}

\subsection{Lagrange-Legendre basis and Bloch-DBMM matrix elements}\label{App:lagrange-legendre-matrix-elements}

The linear system of Eq.~\eqref{eq:cc_linear_system} is obtained from expanding the scattered wave as in Eq.~\eqref{eq:cc_wave_expansion} and projecting Eq.~\eqref{eq:scattered_eq_bloch} onto $f_i(r)$. This subsection collects the definitions and matrix elements it requires.

The Lagrange-Legendre functions are conventionally defined on the
interval $x\in(0,1)$,
\begin{equation}\label{eq:lagrange_legendre_polynomial}
    \hat f_j(x) = (-1)^{N-j}\sqrt{\frac{1-x_j}{x_j}}\,
    \frac{x\,P_N(2x-1)}{x-x_j},
\end{equation}
where the mesh points $x_j$ are the zeros of the shifted Legendre polynomial,
$P_N(2x_j-1)=0$, and $N$ is the number of basis functions per channel. The
associated Gauss--Legendre weights are
\begin{equation}\label{eq:gauss_weights}
    \lambda_j = \frac{1}{4x_j(1-x_j)\left[P_N'(2x_j-1)\right]^2}.
\end{equation}
Quantities are carried to the physical interval $(0,a)$ through $x=r/a$, under
which $f_i(r) = \hat f_i(x)/\sqrt{a}$ and $\delta(r-a) = \delta(x-1)/a$. The Lagrange-Legendre functions vanish at the origin, so Eq.~\eqref{eq:bc_origin} is satisfied by
construction. The basis functions also obey the Lagrange property
\begin{equation}\label{eq:lagrange_property}
    \hat f_i(x_j) = \lambda_i^{-1/2}\,\delta_{ij},
\end{equation}
and their boundary values can be written
\begin{equation}\label{eq:lagrange_boundary_value}
    \hat f_j(1) = \frac{(-1)^{N-j}}{\sqrt{x_j(1-x_j)}}.
\end{equation}

Three matrix elements enter the Bloch-DBMM Hamiltonian. The kinetic operator
$T=-d^2/dr^2$ combined with the diagonal Bloch operator $\mathcal{L}(0)$ gives
the standard R-matrix kinetic block \cite{baye2015lagrange}, for $i\neq j$,
\begin{equation}\label{eq:kinetic_offdiag}
\begin{aligned}
    &\langle \hat f_i | T + \mathcal{L}(0) | \hat f_j \rangle
    = \frac{(-1)^{i-j}}
    {a^2\left[x_i x_j (1-x_i)(1-x_j)\right]^{1/2}} \\
    &\times \left[ N^2 + N + 1
    + \frac{x_i + x_j - 2 x_i x_j}{(x_i - x_j)^2}
    - \frac{1}{1-x_i} - \frac{1}{1-x_j} \right],
\end{aligned}
\end{equation}
and for $i=j$,
\begin{equation}\label{eq:kinetic_diag}
    \langle \hat f_i | T + \mathcal{L}(0) | \hat f_i \rangle
    = \frac{(4N^2 + 4N + 3)\,x_i(1-x_i) - 6x_i + 1}
    {3a^2 x_i^2 (1-x_i)^2}.
\end{equation}
The remaining part of the generalized Bloch operator, $-\delta(r-a)B_{\alpha\gamma}/r$, carries the off-diagonal channel couplings and enters Eq.~(\ref{eq:cc_hamiltonian_matrix}) with a minus sign. Its matrix element follows from the boundary values of Eq.~\eqref{eq:lagrange_boundary_value},
\begin{equation}\label{eq:cc_bloch_surface_element}
    \Big\langle \hat f_i \Big| \delta(r-a)\frac{B_{\alpha\gamma}}{r}
    \Big| \hat f_j \Big\rangle
    = \frac{1}{a}\,[\boldsymbol{\gamma}_s]_{\alpha\gamma}\,
    \hat f_i(1)\,\hat f_j(1).
\end{equation}
Note that local potentials contribute only through their values at the mesh points $r_i=ax_i$, by Eq.~\eqref{eq:lagrange_property}.

Assembling these, the Bloch-DBMM coupled Hamiltonian is written \begin{equation}\label{eq:cc_hamiltonian_matrix}
\begin{aligned}
    \mathcal{H}_{\alpha i,\,\gamma j} = {}&
    \delta_{\alpha\gamma}\Bigg[
    \Braket{ \hat f_i | T + \mathcal{L}(0) | \hat f_j } \\
    &\qquad\quad + \bigg( \frac{\ell_\alpha(\ell_\alpha+1)}{r_i^2}
    - k_\alpha^2 \bigg)\delta_{ij} \Bigg] \\
    & + U_{\alpha\gamma}(r_i)\,\delta_{ij}
    - \frac{1}{a}\,[\boldsymbol{\gamma}_s]_{\alpha\gamma}\,
    \hat f_i(1)\,\hat f_j(1).
\end{aligned}
\end{equation}
Similarly, using Eq.~\eqref{eq:lagrange_property}, the source term can be expressed as:
\begin{equation}\label{eq:cc_source}
    b^{(\beta_0)}_{\alpha i} = -\sqrt{a\lambda_i}
    \sum_{\gamma=1}^{N_{ch}} U^{\mathrm{sr}}_{\alpha\gamma}(r_i)\,
    \bar F^{(\beta_0)}_{\gamma}(r_i).
\end{equation}

\subsection{Symmetrization of the Hamiltonian via the generalized Bloch operator}

In Eq.~\eqref{eq:scattered_eq_bloch}, the generalized Bloch operator restores a symmetric Hamiltonian if and only if the boundary matrix $\boldsymbol{\gamma}_s$ is symmetric. The proof is short. Let $\bar{\mathbf H}^+$ be the $N_{ch}\times N_{ch}$ matrix with elements $\bar H^+_{\alpha\gamma}$. Equation~(\ref{eq:coupled_F}) reads $\bar{\mathbf H}^{+\prime\prime}=\boldsymbol{Q}\bar{\mathbf H}^{+}$ with $Q_{\alpha\gamma}=[\ell_\alpha(\ell_\alpha+1)/r^2-k_\alpha^2]\delta_{\alpha\gamma}+U^C_{\alpha\gamma}$, which is symmetric when $U^C_{\alpha\gamma}=U^C_{\gamma\alpha}$. The condition holds here: in the convention of Ref.~\cite{thompson2009nuclear} the couplings carry a factor $i^{\ell_\gamma-\ell_\alpha}$, which is real and symmetric because all channels of a $J^\pi$ block built on the $K^\pi=0^+$ band have orbital angular momenta of the same parity. The Wronskian matrix
\begin{equation}\label{eq:wronskian_H}
    \boldsymbol{W}(r)=\big(\bar{\mathbf H}^{+}\big)^T\bar{\mathbf H}^{+\prime}
    -\big(\bar{\mathbf H}^{+\prime}\big)^T\bar{\mathbf H}^{+}
\end{equation}
then satisfies $\boldsymbol{W}'=(\bar{\mathbf H}^{+})^T(\boldsymbol{Q}-\boldsymbol{Q}^T)\bar{\mathbf H}^{+}=0$, so it does not depend on $r$. As $r\to\infty$, the off-diagonal parts of $\bar{\mathbf H}^+$ and of its derivative vanish, so the off-diagonal elements of $\boldsymbol W$ tend to zero, while each diagonal element tends to $H^+_{\ell_\alpha}H^{+\prime}_{\ell_\alpha}-H^{+\prime}_{\ell_\alpha}H^+_{\ell_\alpha}=0$. Hence $\boldsymbol W=0$ at every $r$. Multiplying Eq.~(\ref{eq:wronskian_H}) from the left by $[(\bar{\mathbf H}^{+})^T]^{-1}$ and from the right by $(\bar{\mathbf H}^{+})^{-1}$ gives
\begin{equation}
    \boldsymbol{\gamma}_s=\bar{\mathbf H}^{+\prime}(a)\big[\bar{\mathbf H}^{+}(a)\big]^{-1}=\boldsymbol{\gamma}_s^T .
\end{equation}
With symmetric nuclear couplings, the matrix of Eq.~(\ref{eq:cc_hamiltonian_matrix}) is therefore complex symmetric. It is not Hermitian.

\subsection{Relation to the original DBMM}


The Bloch-DBMM solver takes both its name and its boundary condition from Ref.~\cite{dbmm_lei}. We now compare Bloch-DBMM and DBMM formulations.

\textit{(i) Discretization of the radial equation.} Reference~\cite{dbmm_lei} uses collocation. The Schr\"odinger equation is imposed at the $N-1$ interior mesh points through Baye's derivative matrix for $-d^2/dr^2$, which is not symmetric, and the equation at the last mesh point is replaced by the logarithmic-derivative condition. The resulting matrix is complex and non-symmetric. In Bloch-DBMM the same condition enters the weak form of the equation through the surface term of the Bloch operator, Eq.~(\ref{eq:scattered_eq_bloch}). All $N$ mesh points then carry the projected equation, the kinetic block is the symmetric $\langle\hat f_i|T+\mathcal{L}(0)|\hat f_j\rangle$ of the modern R-matrix method, and the boundary condition appears as the term $-a^{-1}[\boldsymbol\gamma_s]_{\alpha\gamma}\hat f_i(1)\hat f_j(1)$. This term is rank one in the mesh indices and couples every mesh point to the surface, since $\hat f_j(1)\neq0$ for all $j$, Eq.~(\ref{eq:lagrange_boundary_value}). Since $\boldsymbol\gamma_s$ is also symmetric, as shown above, the assembled matrix is complex symmetric. The factor of about $1.5$ to $2$ in basis size for the convergence of the cross sections estimated in Ref.~\cite{dbmm_lei} for $p+{}^{12}$C relative to the R-matrix method therefore arises from replacing the last mesh equation by the boundary condition, rather than from the outgoing-wave boundary condition itself. With the weak form, Bloch-DBMM converges at least as fast as the R-matrix method in the cases tested in Appendix~\ref{app:single_channel_scattering_details}.

\textit{(ii) Channel coupling at the boundary.} Reference~\cite{dbmm_lei} decomposes the wave function using the uncoupled regular Coulomb function $\delta_{\alpha\beta_0}F_{\ell_{\beta_0}}$ and imposes a channel-diagonal condition with $\gamma_{s,\alpha}=k_\alpha H^{+\prime}_{\ell_\alpha}/H^+_{\ell_\alpha}$. In DBMM, the source term in channels $\alpha\neq\beta_0$ contains the full coupling $U_{\alpha\beta_0}$, including the Coulomb multipoles that fall off as $r^{-(\omega+1)}$. Because these multipoles decay only as powers of $r$, for a target with a deformed charge distribution this pushes the matching radius required to solve the scattering wave linear system towards $r_\text{asym}=300$~fm. The current implementation of our new Bloch-DBMM uses coupled-channels Coulomb functions $\bar F^{(\beta_0)}_\alpha$ and $\bar{\mathbf H}^+$. It confines the source to the nuclear region and replaces the scalars $\gamma_{s,\alpha}$ by the matrix $\boldsymbol\gamma_s$, so the matching can be done at $a=18$~fm. For neutrons the two decompositions coincide. For charged projectiles without Coulomb deformation they coincide provided the same diagonal Coulomb potential generates the incoming wave. 


\subsection{Formulation of the coupled-channels empirical interpolation method}

In the emulator work by Lei \cite{emulator_dbmm_lei}, the computational bottleneck
for a system coupling many channels was identified, not as the time it takes to
solve the reduced basis equations, but rather the time it takes to construct the
coupling potential and subsequently project the high-fidelity space of
Lagrange--Legendre basis functions to the reduced coordinate space. The Empirical
Interpolation Method allows us to move this computational burden to the offline
stage of the calculation. To see how, we first examine the structure of the
coupling potential.

The most general axially-symmetric coupling potential
$U(\boldsymbol{r}, \boldsymbol{\xi};\vars)$ can be decomposed in multipoles as a
sum of a scalar radial function, a tensor operator, and a spherical harmonic
\cite{thompson2009nuclear}:
\begin{equation}\label{eq:tensor_decomp}
    U(\boldsymbol{r}, \boldsymbol{\xi};\vars) = 
    \sum_{\omega=0}^{N_m} \sum_{m_\omega=-\omega}^{+\omega}
    f^\omega(r;\vars)\,\mathcal{D}_\omega^{m_\omega}(\boldsymbol{\xi})\,
    \mathcal{Y}_{\omega}^{m_\omega}(\boldsymbol{\hat{r}}).
\end{equation}
The first factor $f^\omega(r;\vars)$ is a scalar function of the radial distance
$r$. The second, $\mathcal{D}_\omega^{m_\omega}(\boldsymbol{\xi})$, is the nuclear
transition operator; for a rotational nucleus, for example, it is the rotation
matrix transforming from the lab frame to the target body-fixed frame
characterized by the three Euler angles $\boldsymbol{\xi}$. The third is a standard
spherical harmonic. The number of multipoles $N_m$ retained in the expansion
depends on the choice and degree of the deformation. The radial scalar function is
obtained by integrating out the angular dependence of the deformed potential,
projecting onto a given multipolarity $\omega$,
\begin{equation}\label{eq:scalar_function_int}
    f^\omega(r;\vars) = 2\pi\int_0^{\pi}
    U(\boldsymbol{r},\boldsymbol{\xi};\vars)\,\mathcal{Y}_{\omega}(\boldsymbol{\hat{r}}')\,
    \sin\theta'\,d\theta'.
\end{equation}
Projecting Eq.~\eqref{eq:tensor_decomp} onto the channel states of the $J^\pi$
block gives the coupling matrix elements between channels $\alpha$ and $\gamma$,
\begin{equation}\label{eq:geometric_coupling}
    U_{\alpha\gamma}(r;\vars)
    = \sum_{\omega=0}^{N_m}
    \mathcal{V}_{\alpha\gamma}^\omega\, f^\omega(r;\vars),
\end{equation}
where $\mathcal{V}_{\alpha\gamma}^\omega$ are the \textit{geometric} coupling
strengths of multipolarity $\omega$, obtained from the angular-momentum algebra of
the chosen coupling scheme.

The decomposition in Eq.~\eqref{eq:geometric_coupling} separates the parameter
$\vars$ and radial $r$ dependence, carried by $f^\omega(r;\vars)$, from the
\textit{geometric} part of the coupling potential, which depends on
$\boldsymbol{\xi}$ and $\boldsymbol{\hat{r}}$ and is fixed for a given
projectile--target system and choice of couplings. The radial function $f_\omega$
is not an affine function of the parameters $\vars$, which is what makes the
projection onto the reduced basis costly to repeat for each new $\vars$. The
Empirical Interpolation Method
\cite{barrault2004empirical, grepl2007efficient, odell2024rose, wavefunction-based-catacorarios}
removes this difficulty by writing each $f^\omega$ as a sum of fixed basis
functions weighted by parameter-dependent coefficients. In this form the radial
integrals that perform the projection no longer depend on $\vars$, so they can be
computed once and stored. In previous work
\cite{wavefunction-based-catacorarios}, we assumed an analytic, small-deformation
limit of the radial scalar functions $f_\omega(r;\vars)$. We now extend this
formalism to any coupling potential of any model which allows the decomposition of
Eq.~\eqref{eq:tensor_decomp}. The implementation is straightforward provided the
angular integral of Eq.~\eqref{eq:scalar_function_int} can be evaluated
efficiently. We build this representation as follows.

We start from a principal-component analysis of the sampled form factors. We
evaluate the radial function $f^\omega(r;\vars)$ at $n_U$ sets of training
parameters $\{\vars_j\}_{j=1}^{n_U}$, on a dense radial grid $\{r_q\}_{q=1}^{Q}$. The angular integration of Eq.~\eqref{eq:scalar_function_int} is
computed using Gauss--Legendre quadrature. We then collect the results as the
columns of an interaction snapshot matrix $\boldsymbol{\Phi}^\omega$, with
$[\boldsymbol{\Phi}^\omega]_{qj} = f^\omega(r_q;\vars_j)$. The principal components
follow from the singular value decomposition
\begin{equation}\label{eq:eim_pca}
    \boldsymbol{\Phi}^\omega = \boldsymbol{A}^\omega\,\boldsymbol{\Sigma}^\omega\,
    (\boldsymbol{B}^{ \omega})^\dagger,
\end{equation}
where the diagonal entries $\sigma^{\omega}_{m}$ of $\boldsymbol{\Sigma}^\omega$
are the singular values, ordered by size. We keep the $N_U$ leading columns of
$\boldsymbol{A}^\omega$ as the interaction basis,
\begin{equation}\label{eq:eim_basis}
    u^{\omega}_{m}(r_q) = \big[\boldsymbol{A}^\omega\big]_{qm},
    \qquad m = 1,\dots,N_U ,
\end{equation}
and approximate the form factor at any parameter set $\vars$ as a linear
combination of these basis functions,
\begin{equation}\label{eq:eim_expand}
    f^\omega(r;\vars) \approx
    \sum_{m=1}^{N_U} e^{\omega}_{m}(\vars)\,u^{\omega}_{m}(r),
\end{equation}
with parameter-dependent coefficients $e^{\omega}_{m}(\vars)$. The singular values
measure how much each basis function contributes, and their rapid decay is what
lets us truncate to a small number $N_U \ll n_U$ of terms. In practice, $N_U$ is
fixed by a tolerance $\epsilon$: we retain all basis functions whose singular value
satisfies
\begin{equation}\label{eq:eim_tolerance}
    \sigma^{\omega}_{m} > \epsilon\, \sigma^{\omega}_{1},
    \qquad m = 1,2,\dots,N_U .
\end{equation}

The coefficients are fixed by requiring that Eq.~\eqref{eq:eim_expand} reproduce
the exact form factor at $N_U$ collocation points $\{r_p\}_{p=1}^{N_U}$. These are
selected from the dense grid by the MaxVol algorithm \cite{Goreinov_2010}, which
maximizes the determinant of the resulting submatrix of $u^{\omega}_{m}(r_q)$;
further details can be found in \cite{odell2024rose}. The interpolation condition
then reduces to the linear system
\begin{equation}\label{eq:EIM_coefficients}
    \boldsymbol{e}^\omega(\vars)
    = \big(\boldsymbol{M}^\omega\big)^{-1}\cdot
    \boldsymbol{\phi}^\omega(\vars),
\end{equation}
with the $N_U\times N_U$ collocation matrix
\begin{equation}\label{eq:EIM_matrix_form_a}
    \boldsymbol{M}^\omega = \begin{bmatrix}
        u^{\omega}_{1}(r_1) & u^{\omega}_{2}(r_1) & \dots & u^{\omega}_{N_U}(r_1) \\
        u^{\omega}_{1}(r_2) & u^{\omega}_{2}(r_2) & \dots & u^{\omega}_{N_U}(r_2) \\
        \vdots & \vdots & \ddots & \vdots \\
        u^{\omega}_{1}(r_{N_U}) & u^{\omega}_{2}(r_{N_U}) & \dots &
        u^{\omega}_{N_U}(r_{N_U})
    \end{bmatrix},
\end{equation}
and the vector of exact evaluations at those points,
\begin{equation}\label{eq:EIM_matrix_form}
    \boldsymbol{\phi}^\omega(\vars) = \begin{bmatrix}
        f^\omega(r_{1};\vars) \\
        f^\omega(r_{2};\vars) \\
        \vdots \\
        f^\omega(r_{N_U};\vars)
    \end{bmatrix}.
\end{equation}
Since $\boldsymbol{M}^\omega$ does not depend on $\vars$, it is inverted once in the
offline stage; for each new parameter set the online work is the evaluation of
$f^\omega$ at the $N_U$ collocation points and the matrix--vector product of
Eq.~\eqref{eq:EIM_coefficients}.

The expansion of Eq.~\eqref{eq:eim_expand} is required at the Lagrange--Legendre
mesh points $r_k = a x_k$, whereas the basis was constructed on the dense grid
$\{r_q\}$. Rather than storing and interpolating the singular vectors, we
re-evaluate the basis directly at the Lagrange points. This works because of the
structure of Eq.~\eqref{eq:eim_pca}: the left singular vectors
$\boldsymbol{A}^\omega$ carry the radial dependence of the basis, while the right
singular vectors $\boldsymbol{B}^\omega$ and the singular values
$\boldsymbol{\Sigma}^\omega$ depend only on the training parameters
$\{\vars_j\}$, not on the radial grid. Writing out a single column of
Eq.~\eqref{eq:eim_pca} expresses each basis function as a weighted sum of the
snapshots,
\begin{equation}\label{eq:eim_left_vector}
    u^{\omega}_{m}(r_q)
    = \frac{1}{\sigma^{\omega}_{m}}
      \sum_{j=1}^{n_U} f^\omega(r_q;\vars_j)\,\big[\boldsymbol{B}^\omega\big]_{jm},
\end{equation}
in which the only radius-dependent quantity is the snapshot value
$f^\omega(r_q;\vars_j)$. The weights $[\boldsymbol{B}^\omega]_{jm}$ and the factor
$1/\sigma^{\omega}_{m}$ are unchanged no matter where the form factor is evaluated.
We can therefore obtain the basis at the Lagrange points by evaluating the same
training interactions at those points and combining them with the same weights,
\begin{equation}\label{eq:eim_basis_abscissa}
    u^{\omega}_{m}(r_k)
    = \frac{1}{\sigma^{\omega}_{m}}
      \sum_{j=1}^{n_U} f^\omega(r_k;\vars_j)\,\big[\boldsymbol{B}^\omega\big]_{jm} .
\end{equation}
The emulated form factor on the Lagrange mesh then follows from
Eq.~\eqref{eq:eim_expand},
\begin{equation}\label{eq:eim_emulated_abscissa}
    f^\omega(r_k;\vars) \approx
    \sum_{m=1}^{N_U} e^{\omega}_{m}(\vars)\,u^{\omega}_{m}(r_k).
\end{equation}
Constructing the basis on the dense grid, selecting the optimal $N_U$ collocation points
from it, and reconstructing the basis once on the Lagrange mesh
$\{r_k\}_{k=1}^{N}$ makes the online cost independent of the number of Lagrange
mesh points chosen. This is the origin of the computational advantage of our method for
non-affine parameters.

We now assemble the matrices used in Sec.~\ref{sec:formalism}. Throughout this subsection, a matrix in the $(N \cdot N_{ch}) \times (N \cdot N_{ch})$ Bloch-DBMM space contains 4 pairs of indices, with $\alpha,\, \gamma$ running over the
channel index and $k,\, l $ denoting the Lagrange mesh quadrature point.

The geometric coupling strengths of Eq.~\eqref{eq:geometric_coupling} collect into
an $N_{ch}\times N_{ch}$ matrix for each multipolarity,
\begin{equation}\label{eq:geom_matrix}
    \big[\boldsymbol{\mathcal{V}}^{\omega}\big]_{\alpha\gamma}
    = \mathcal{V}_{\alpha\gamma}^\omega,
\end{equation}
fixed for a given projectile--target system and coupling scheme. Because the
coupling potential is local, its radial factor is diagonal on the Lagrange mesh,
\begin{equation}\label{eq:radial_matrix}
    \big[\boldsymbol{f}^{\omega}(\vars)\big]_{kl}
    = f^\omega(r_k;\vars)\,\delta_{kl},
\end{equation}
which follows from the Gauss--Legendre quadrature approximation together with the
Lagrange property of Eq.~\eqref{eq:lagrange_property}. The multipole component of
the potential matrix in Eq.~\eqref{eq:V_multipole_matrix} is therefore
\begin{equation}\label{eq:F_omega_index}
    \big[\boldsymbol{\mathcal{F}}^{\omega}(\vars)\big]_{\alpha k,\,\gamma l}
    = \mathcal{V}_{\alpha\gamma}^\omega\, f^\omega(r_k;\vars)\,\delta_{kl},
\end{equation}
which is the index form of the outer product
$\boldsymbol{\mathcal{F}}^{\omega} = \boldsymbol{\mathcal{V}}^{\omega} \otimes
\boldsymbol{f}^{\omega}$.

Replacing the exact form factor by its EIM expansion,
Eq.~\eqref{eq:eim_emulated_abscissa}, gives the fixed basis matrices of
Eq.~\eqref{eq:eim_matrix_main},
\begin{equation}\label{eq:g_omega_m}
    \big[\boldsymbol{u}^{\omega}_{m}\big]_{kl}
    = u^{\omega}_{m}(r_k)\,\delta_{kl},
    \qquad
    \big[\boldsymbol{G}^{\omega}_{m}\big]_{\alpha k,\,\gamma l}
    = \mathcal{V}_{\alpha\gamma}^\omega\, u^{\omega}_{m}(r_k)\,\delta_{kl},
\end{equation}
with coefficients $e^{\omega}_{m}(\vars)$ given by
Eq.~\eqref{eq:EIM_coefficients}. Since the $u^{\omega}_{m}(r_k)$ are obtained once
from Eq.~\eqref{eq:eim_basis_abscissa} and the geometric factors carry no parameter
dependence, the matrices $\boldsymbol{G}^{\omega}_{m}$ are built and projected onto
the reduced basis entirely in the offline stage.

\subsection{Benchmark to \texttt{Frescox}}

We collect here the quantitative comparison between the Bloch-DBMM solver and
\texttt{Frescox} for the two systems and energies considered in
Sec.~\ref{sec:Implementation}. All calculations use the RIPL-3 parameters, a
matching radius $a=18$~fm, $J_{\mathrm{max}}=20.5$, and $N=60$ Lagrange--Legendre
basis states per channel.

\begin{figure}
    \centering
    \includegraphics[width=1.0\linewidth]{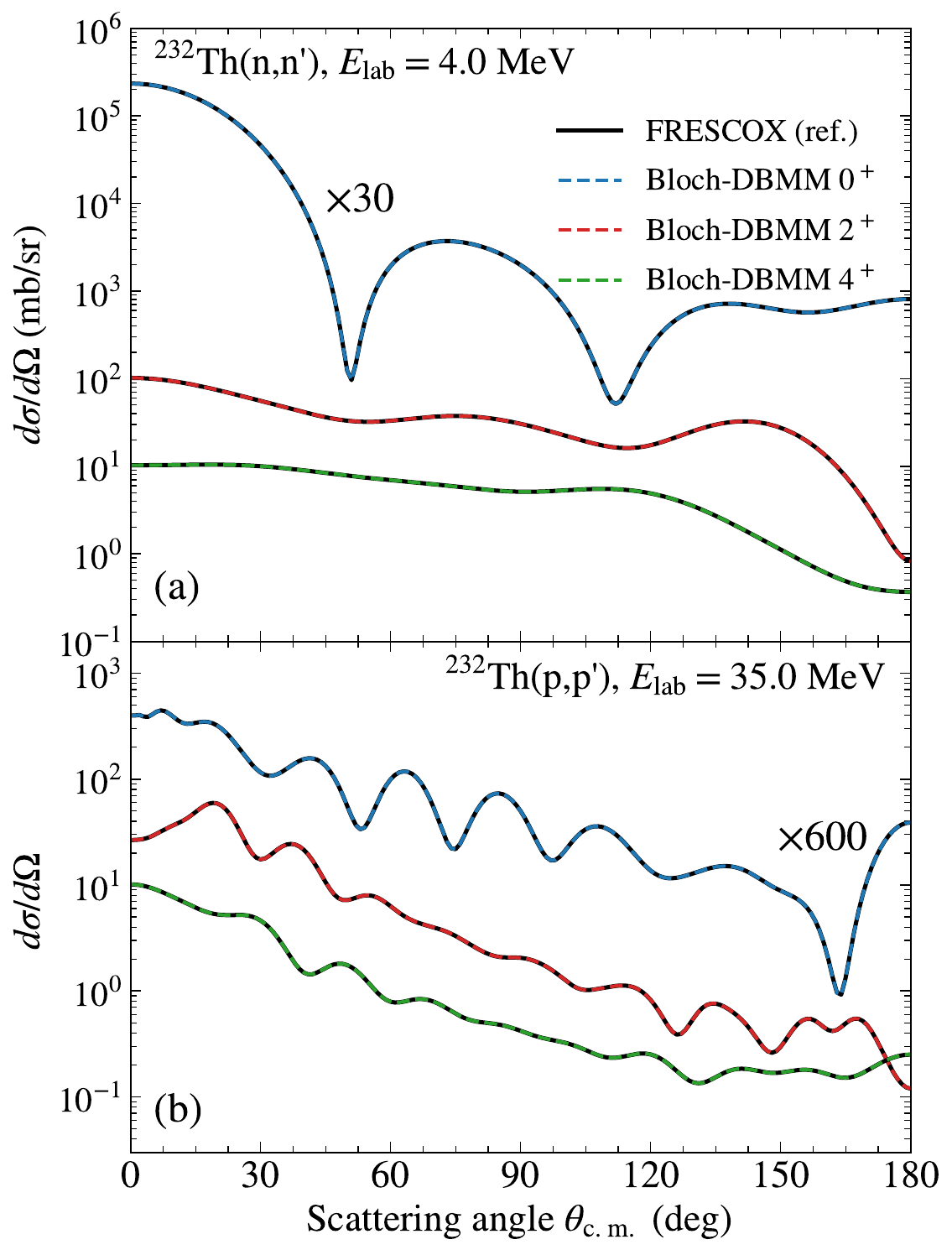}
    \caption{\label{fig:frescox_vs_bloch}Differential cross sections for
$^{232}$Th$(n,n')$ at $E_\mathrm{lab}=4$ MeV (a) and $^{232}$Th$(p,p')$ at
$E_\mathrm{lab}=35$ MeV (b), for the first three members of the ground-state
rotational band. Solid black lines denote the \texttt{Frescox} reference;
dashed coloured lines the Bloch-DBMM result for the $0^+$ (blue), $2^+$ (red),
and $4^+$ (green) states. In panel (a) the $0^+$ cross section is scaled by a
factor of 30; in panel (b) it is shown as a ratio to the Rutherford cross
section and scaled by 600. All remaining cross sections are in mb/sr. Both
calculations use $a=18$ fm, $J_\mathrm{max}=20.5$, and $N=60$
Lagrange--Legendre basis states per channel.}
\end{figure}

In Fig.~\ref{fig:frescox_vs_bloch} we plot the differential cross sections for the first three members of the rotational band calculated with Bloch-DBMM and compare them against the equivalent \texttt{Frescox} calculation, using the OMP parameters obtained from RIPL-3. Panel (a) shows the low-energy neutron case, with the $0^+$ cross section scaled by a factor of 30 for clarity. Panel (b) shows the high-energy proton case, which tests the accuracy of the nuclear part of our calculation; there the $0^+$ state is shown as a ratio to the Rutherford cross section and scaled by 600. All remaining cross sections are in mb/sr.

Table~\ref{tab:frescox} gives the median relative error in the differential cross
section for each member of the ground-state rotational band. For the neutron case
the agreement is at the $10^{-5}$ level and is uniform across the band. For
protons the agreement is comparable at $4$~MeV, where the cross section is
dominated by the Coulomb interaction, and is at the $10^{-4}$ level at $35$~MeV.
The larger error at $35$~MeV comes from the long range of the coupling potential
for charged projectiles, which does not vanish within $a=18$~fm and therefore
requires more basis states than the neutron case. Increasing $N$ or $a$ reduces
this difference, at the corresponding computational cost.

\begin{table}
\begin{ruledtabular}
\begin{tabular}{ccc}
State & 4 MeV & 35 MeV \\
\colrule
\multicolumn{3}{c}{$^{232}$Th$(n,n')$} \\
\colrule
$0^+$ & $7.214\times10^{-5}$ & $7.082\times10^{-5}$ \\
$2^+$ & $6.923\times10^{-5}$ & $7.689\times10^{-5}$ \\
$4^+$ & $6.069\times10^{-5}$ & $7.099\times10^{-5}$ \\
$6^+$ & $9.788\times10^{-5}$ & $4.898\times10^{-5}$ \\
$8^+$ & $1.144\times10^{-4}$ & $4.731\times10^{-5}$ \\
All   & $7.961\times10^{-5}$ & $6.170\times10^{-5}$ \\
\colrule
\multicolumn{3}{c}{$^{232}$Th$(p,p')$} \\
\colrule
$0^+$ & $1.879\times10^{-5}$ & $1.997\times10^{-4}$ \\
$2^+$ & $3.782\times10^{-5}$ & $4.326\times10^{-4}$ \\
$4^+$ & $7.543\times10^{-5}$ & $1.883\times10^{-4}$ \\
$6^+$ & $7.123\times10^{-5}$ & $1.583\times10^{-4}$ \\
$8^+$ & $3.330\times10^{-4}$ & $2.302\times10^{-4}$ \\
All   & $4.312\times10^{-5}$ & $2.212\times10^{-4}$ \\
\end{tabular}
\end{ruledtabular}
\caption{\label{tab:frescox}Median relative error in the differential cross
section $d\sigma/d\Omega$, with respect to the {\sc Frescox} reference, for each
member of the $^{232}$Th ground-state rotational band, for neutron and proton
inelastic scattering at 4 and 35~MeV computed with the Bloch--DBMM method. The
``All'' rows give the median over all states for each reaction. All
calculations use matching radius $a=18$~fm, $J_{\mathrm{max}}=20.5$, and $N=60$
Lagrange--Legendre basis states per channel.}
\end{table}

Table~\ref{tab:tcoef} gives the relative error in the spin-averaged transmission
coefficients $\bar{T}_\ell$ for the first five orbital angular momenta, together
with that in the total reaction cross section $\sigma_R$. These are the quantities
that feed statistical reaction codes, and the agreement here is one to two orders
of magnitude tighter than for the differential cross sections, reaching
$10^{-7}$--$10^{-6}$ for neutrons and $10^{-6}$--$10^{-5}$ for protons at $35$~MeV.
This is expected, since $\bar{T}_\ell$ and $\sigma_R$ depend only on the moduli of
the $S$ matrix elements, whereas the angular distribution is a coherent sum in
which phase errors also contribute. At $4$~MeV the proton transmission
coefficients are suppressed to order $10^{-6}$ by the Coulomb barrier, and we do
not quote their relative error.

\begin{table}
\begin{ruledtabular}
\begin{tabular}{ccc}
$\bar{T}_\ell$ & 4 MeV & 35 MeV \\
\colrule
\multicolumn{3}{c}{$^{232}$Th$(n,n')$} \\
\colrule
1 & $2.95\times10^{-6}$ & $7.88\times10^{-7}$ \\
2 & $7.74\times10^{-7}$ & $3.60\times10^{-7}$ \\
3 & $2.38\times10^{-7}$ & $1.22\times10^{-7}$ \\
4 & $1.12\times10^{-6}$ & $1.80\times10^{-7}$ \\
5 & $1.97\times10^{-7}$ & $1.21\times10^{-7}$ \\
\colrule
$\sigma_R$ & $5.3\times10^{-7}$ & $1.0\times10^{-7}$ \\
\colrule
\multicolumn{3}{c}{$^{232}$Th$(p,p')$} \\
\colrule
1 & NA & $3.93\times10^{-6}$ \\
2 & NA & $5.07\times10^{-6}$ \\
3 & NA & $3.09\times10^{-6}$ \\
4 & NA & $1.44\times10^{-5}$ \\
5 & NA & $9.19\times10^{-6}$ \\
\colrule
$\sigma_R$ & NA & $3.0\times10^{-6}$ \\
\end{tabular}
\end{ruledtabular}
\caption{\label{tab:tcoef}Relative error in the averaged transmission
coefficient $\bar{T}_\ell=[T_{\ell,\ell+1/2}+T_{\ell,\ell-1/2}]/2$
, computed
with the Bloch--DBMM method relative to the {\sc Frescox} reference, for the
first five orbital angular momenta $\ell$ in neutron and proton inelastic
scattering off $^{232}$Th at 4 and 35~MeV. The $\sigma_R$ row of each block
gives the relative error in the total reaction cross section. The proton transmission coefficients are suppressed to the order of $10^{-6}$ at 4~MeV by the Coulomb
barrier, so we do not include their relative error. All calculations use matching radius $a=18$~fm, $J_{\mathrm{max}}=20.5$,
and $N=60$ Lagrange--Legendre basis states per channel.}
\end{table}


\subsection{Construction of the coupled Coulomb incoming wave}

Integrating Eq.~\eqref{eq:coupled_F} outward from the origin with a regular Coulomb seed in channel $j$ gives one regular solution per channel. Collect them as the columns of an $N_{ch}\times N_{ch}$ fundamental solution matrix $\mathbf{Y}(r)$; every solution regular at the origin is a linear combination $\mathbf{Y}(r)\mathbf{z}$ with constant coefficients $\mathbf{z}$. For $r>R_c$, where $U^C$ is point-like, each column is a superposition of the coupled outgoing and incoming
Coulomb functions,
\begin{equation}\label{eq:Y_expansion}
    \mathbf{Y}(r) = \bar{\mathbf{H}}^+(r)\,\mathbf{\Xi}^+
                  + \bar{\mathbf{H}}^-(r)\,\mathbf{\Xi}^-, \qquad r\ge a,
\end{equation}
where $\bar{\mathbf{H}}^\pm$ are the solutions of Eq.~\eqref{eq:coupled_F} that tend to $\delta_{\alpha\gamma}H^\pm_{\ell_\alpha}(\eta_\alpha,k_\alpha r)$, obtained by the inward Coulomb propagation of Sec.~\ref{sec:Implementation} \cite{fresco}; for open channels and real $U^C$, $\bar{\mathbf{H}}^-=(\bar{\mathbf{H}}^+)^*$.
The constant matrices $\mathbf{\Xi}^\pm$ follow by imposing Eq.~\eqref{eq:Y_expansion} and its derivative at $r=a$, a linear system of dimension $2N_{ch}$. The matrix $\mathbf{\Xi}^-$ gives the incoming content of each regular solution and
$\mathbf{\Xi}^+$ its outgoing content.

Since $F_\ell=(H^+_\ell-H^-_\ell)/2i$, the free Coulomb wave carries incoming amplitude $i/2$. The normalization of Eq.~\eqref{eq:Fbar_asym} therefore requires
\begin{equation}\label{eq:d_condition}
    \big[\mathbf{\Xi}^-\mathbf{z}^{(\beta_0)}\big]_\alpha = \tfrac{i}{2}\,\delta_{\alpha\beta_0}.
\end{equation}
Solving this $N_{ch}\times N_{ch}$ system fixes $\mathbf{z}^{(\beta_0)}$, and hence $\bar{F}^{(\beta_0)}=\mathbf{Y}\mathbf{z}^{(\beta_0)}$. Its outgoing content is then \begin{equation}\label{eq:c_from_M}
    \big[\mathbf{\Xi}^+\mathbf{z}^{(\beta_0)}\big]_\alpha = -\tfrac{i}{2}\,\delta_{\alpha\beta_0} + t^C_{\alpha\beta_0},
\end{equation}
where the first term is the outgoing part of $\delta_{\alpha\beta_0}F_{\ell_{\beta_0}}$ and $t^C_{\alpha\beta_0}$ is the Coulomb-excitation amplitude in Eq.~\eqref{eq:SC}. Regularity at the origin holds by construction, so no irregular admixture is introduced at $r=a$.

\subsection{The Coulomb deformation}

The Coulomb coupling potential $U^{C}_{\alpha\gamma}$ of Eq.~\eqref{eq:coupled_F} is generated by the deformation of the target charge distribution. Its strength is set by the electromagnetic coupling strengths $M(E\omega)$, which carry units of $e\,\cdot \mathrm{fm}^{\omega}$ and are defined from the reduced matrix element of Alder and Winther \cite{alder_winther_1966, alder_winther_1975} as
\begin{equation}\label{eq:MEw_reduced}
    M(E\omega) = i^{\,I-I'+|I-I'|}\,\Braket{I' \| E\omega \| I},
\end{equation}
the phase factor here is chosen to keep the expression real-valued. Written this way, the coupling strength is fixed directly by the observable transition rate,
\begin{equation}\label{eq:MEw_from_BEw}
    M(E\omega) = \pm \sqrt{(2I+1)\,B(E\omega, I\rightarrow I')},
\end{equation}
with no model-dependent parameters entering beyond the sign, which follows the convention of \cite{fresco}.

Within the rotational model, $M(E\omega)$ is related to the Coulomb deformation lengths $\delta^{C}_\omega$ by 
\begin{equation}\label{eq:MEw_from_delta}
\begin{aligned}
       M(E\omega, I\rightarrow I')
    = & \frac{3Z\,\delta^{C}_\omega R_c^{\,\omega-1}}{4\pi}\,
    i^{\,I-I'+|I-I'|} \\
    &\times \sqrt{2I+1}\;\Braket{IK\,\omega 0 | I'K},
\end{aligned}
\end{equation}
for transitions between states of spin $I$ and $I'$ in a rotational band of projection $K$ in a nucleus of charge $Z$. It is at this point that the model-dependent Coulomb radius $R_c$ enters. For the $K=0$ bands considered in this work, $M(E\omega, 0\rightarrow I) = M(E\omega, I\rightarrow 0)$ and both carry the same sign as $\delta^{C}_\omega$. We note that a single deformation length gives a different matrix element for each transition within the band, since the angular-momentum coupling in Eq.~\eqref{eq:MEw_from_delta} depends on the spins of the two states connected.

The radial form factors for $\omega\ge2$ follow from the multipole expansion of $|\boldsymbol{r}-\boldsymbol{r}'|^{-1}$,
\begin{equation}\label{eq:coulomb_form_factor}
    f^{C}_\omega(r) = \frac{\sqrt{4\pi}\,e^2}{2\omega+1}\,M(E\omega)
    \begin{cases}
        r^{\omega}/R_c^{\,2\omega+1}, & r \le R_c, \\[4pt]
        1/r^{\,\omega+1}, & r > R_c,
    \end{cases}
\end{equation}
where the factor $\delta^{C}_\omega R_c^{\,\omega-1}$ is already contained in $M(E\omega)$ through Eq.~\eqref{eq:MEw_from_delta}. These form factors are multiplied by the geometric coupling matrix elements $\mathcal{V}_{\omega\alpha\gamma}$ of Eq.~\eqref{eq:geometric_coupling} and by the charge of the projectile to give the multipole part of the full interaction $U_{\alpha\gamma}$. The same terms are carried over unchanged into $U^{C}_{\alpha\gamma}$, so they cancel in $U^{\mathrm{sr}}$. The interior branch $r^{\omega}/R_c^{2\omega+1}$ is the potential inside a uniformly charged sphere with a deformed surface.

The monopole is treated differently. The diagonal Coulomb term of $U$ is the potential of a uniformly charged sphere of radius $R_c$,
\begin{equation}\label{eq:coulomb_sphere}
    V^C_{\mathrm{sph}}(r) = Z_pZ_te^2
    \begin{cases}
        \big(3-r^2/R_c^2\big)/2R_c, & r\le R_c,\\[4pt]
        1/r, & r>R_c,
    \end{cases}
\end{equation}
while $U^C$ contains only its point-charge limit $Z_pZ_te^2/r$, which enters the interior integration through $2\eta_\alpha k_\alpha/r$ and the exterior Coulomb functions through $\eta_\alpha$. The difference $\Delta V_C=V_{\mathrm{sph}}-Z_pZ_te^2/r$ vanishes for $r>R_c$ and is the only Coulomb term left in $U^{\mathrm{sr}}$.


In this work both the Coulomb radius $R_c$ and the Coulomb deformation lengths $\delta^{C}_\omega$ are held fixed at their RIPL-3 values and are not included in $\vars$. The coupling strengths $M(E\omega)$, and with them the Coulomb form factors of Eq.~\eqref{eq:coulomb_form_factor}, are therefore the same for every parameter set, so the coupled-channels Coulomb functions and the boundary term $B_{\alpha\gamma}$ need be computed only once. We emphasize that the Coulomb deformation lengths $\delta^{C}_\omega$ are distinct from the nuclear deformation lengths $\delta_\omega$ of Eq.~\eqref{eq:radial_form_WS}, which are sampled across $\mathcal{T}$. It is common in optical model calibrations to treat $\delta^{C}_\omega$ separately from the nuclear deformation lengths $\delta_\omega$ of Eq.~\eqref{eq:radial_form_WS} \cite{CAPOTE_2008}.

\bibliographystyle{apsrev4-1}

\bibliography{gevc} 
\end{document}